\documentclass[10pt]{article}

\usepackage{amsmath}
\usepackage{amssymb}  
\usepackage[T1]{fontenc}
\usepackage{pifont}
\usepackage{pstricks}
\usepackage{amsfonts}
\usepackage{graphicx}
\usepackage{amsmath}
\usepackage{authblk}
\usepackage{pstricks} 
\usepackage{pstricks-add}
\usepackage{caption}
\usepackage{placeins}
\usepackage{pstricks}
\usepackage{pdfpages}
\usepackage{framed}
\usepackage{bigints}
\usepackage{cancel}
\usepackage{bm}
\usepackage{epstopdf}
\usepackage{tikz}
\newcommand*\circled[1]{\tikz[baseline=(char.base)]{
		\node[shape=circle,draw,inner sep=2pt] (char) {#1};}}

\def\rr2dot{\mathop{\bf r}\limits}
\def\x2dot{\mathop{x}\limits}
\def\y2dot{\mathop{y}\limits}

\def\bfy2dot{\mathop{\bf y}\limits}
\def\z2dot{\mathop{z}\limits}

\def\csi2dot{\mathop{\xi}\limits}
\def\et2dot{\mathop{\eta}\limits}
\def\bet2dot{\mathop{\beta}\limits}
\def\t2dot{\mathop{\theta}\limits}
\def\s2dot{\mathop{\sigma}\limits}
\def\d2dot{\mathop{\delta}\limits}
\def\q2dot{\mathop{q}\limits}
\def\l2dot{\mathop{\lambda}\limits}
\def\ps2dot{\mathop{{\cal E}}\limits}
\def\tet2dot{\mathop{\theta}\limits}
\def\bfx2dot{\mathop{\bf x}\limits}
\def\bfy2dot{\mathop{\bf y}\limits}
\def\bfq2dot{\mathop{\bf q}\limits}
\def\bfr2dot{\mathop{\bf r}\limits}
\def\bbfq2dot{\mathop{\bar {\bf q}}\limits}
\def\w2{\mathop{W}\limits}
\def\xgrande2dot{\mathop{\bf X}\limits}
\def\p02dot{\mathop{P}\limits}
\def\a2dot{\mathop{A}\limits}

\newtheorem{teo}{Theorem}
\newtheorem{prop}{Proposition}
\newtheorem{cor}{Corollary}
\newtheorem{propr}{Property}
\newtheorem{rem}{Remark}
\title{On the transpositional relation for nonholonomic systems}
\author{F.~Talamucci}
\affil{{\it DIMAI, Dipartimento di Matematica e Informatica ``Ulisse Dini''},\\
{\it	Universit\`a degli Studi di Firenze, Italy}\\
{\it	e-mail: federico.talamucci@unifi.it}}
\date{}

\begin{document}
	
	\bibliographystyle{plain}
	
	\setcounter{equation}{0}

	\maketitle
	
	\vspace{.5truecm}
	
	
	\vspace{.5truecm}
	
	\noindent
	{\bf Keywords:} 
	
	Nonholonomic mechanical systems -Virtual displacements for nonlinear kinematic constraints - 
	
	${\check {\rm C}}$etaev condition - Transpositional rule, commutation relations.
	
	\vspace{.5truecm}
	
\begin{abstract}
	
	\noindent
	This paper investigates the dynamics of nonholonomic mechanical systems, with a particular focus on the fundamental variational assumptions and the role of the transpositional rule. We analyze how the $\check {\rm C}$etaev condition and the first variation of constraints define compatible virtual displacements for systems subject to kinematic constraints, which can be both linear and nonlinear in generalized velocities. The study meticulously explores the necessary conditions for the commutation relations to hold, clarifying their impact on the consistency of the derived equations of motion. By detailing the interplay between these variational identities and the Lagrangian derivatives of the constraint functions, we shed light on the differences between equations of motion formulated via d'Alembert--Lagrange principle and those obtained from extended time-integral variational principles. This work aims to provide a clearer theoretical framework for understanding and applying these core principles in the complex domain of nonholonomic dynamics.
	
\end{abstract}

\section{Introduction}

\noindent
Describing the dynamics of nonholonomic mechanical systems stands as a persistent and foundational challenge within analytical mechanics. Unlike their holonomic counterparts, which are governed by integrable constraints that allow for an elegant reduction of degrees of freedom and a unified Lagrangian formulation, nonholonomic systems are characterized by kinematic constraints dependent on generalized velocities that cannot be integrated. This inherent non-integrability necessitates a meticulous and often complex re-evaluation of fundamental principles, particularly regarding the definitions of virtual variations and their compatibility with the imposed restrictions on motion. The rigorous formulation of equations of motion for such systems has, for these reasons, historically been a subject of considerable academic debate, leading to the development of diverse and sometimes conflicting theoretical approaches in the literature.

\noindent
This study aims to provide a deeper and more cohesive understanding of these intricate systems by focusing on the transpositional rule. This mathematical identity serves as a crucial conceptual and formal bridge, connecting various conventions and interpretations adopted for handling variations within nonholonomic dynamics. We specifically delve into a detailed analysis of how this rule, alongside key foundational concepts such as ${\check {\rm C}}$etaev's condition (which rigorously defines "ideal" virtual displacements for systems with non-integrable constraints) and the first variation of constraints (a necessary requirement derived from the application of variational principles), fundamentally influences and shapes the formulation of the equations of motion. Our investigation extends to how these principles interact and the implications of their interplay.

\noindent
The paper will meticulously analyze the intricate interaction between these fundamental conditions, a relationship that has often been a source of skepticism and differing viewpoints in the existing scientific literature. A central theme of our exploration will be the necessary conditions for the commutation of variational and time derivative operators. This investigation is of paramount importance because it clarifies that different assumptions regarding these commutation relations are not merely formal details, but rather carry profound repercussions for the consistency and the ultimate physical form of the derived dynamic equations. Our analysis will proceed by thoroughly comparing and contrasting approaches based on the d'Alembert-Lagrange Principle with those derived from generalized time-integral variational principles, including prominent frameworks such as the Hamilton-Suslov Principle and vakonomic mechanics. By highlighting the subtle yet critical distinctions and connections between these methodologies, this work seeks to offer a more unified and rigorous perspective on the dynamics of nonholonomic systems.

\subsection{The physical framework}

\noindent
We consider a mechanical system with generalized coordinates ${\bf q}=(q_1, \dots, q_n)$,  generalized velocities ${\dot {\bf q}}=({\dot q}_1, \dots, {\dot q}_n)$ and subject to the nonholonomic restrictions
\begin{equation}
	\label{vincg}
	g_\nu ({\bf q}, {\dot {\bf q}}, t)=0, \qquad \nu=1, \dots, \kappa<n
\end{equation}
which satisfy the non--singularity condition
\begin{equation}
	\label{gind}
	rank\, \left(\dfrac{\partial (g_1, \dots, g_\kappa)}{\partial ({\dot q}_1, \dots, {\dot q}_n)}\right)=\kappa
\end{equation}
where the quantity in round brackets is the $\kappa\times n$ jacobian matrix with entries $\frac{\partial g_\nu}{\partial {\dot q}_i}$, $\nu=1, \dots, \kappa$, $j=1, \dots, n$.
Each function $g_\nu$ can be linear (as it is assumed in most models in literature) or even nonlinear with respect to the generalized velocities ${\dot q}_i$, $i=1, \dots, n$. In the linear case the functions (\ref{vincg}) are 
\begin{equation}
	\label{glin}
	g_\nu ({\bf q}, {\dot {\bf q}},t)=\sum\limits_{j=1}^n a_{\nu,j}({\bf q,t}){\dot q}_j+b_\nu({\bf q},t)
\end{equation}
and (\ref{gind}) means that the rank of the matrix with entries $(a_{\nu,j})$, $\nu=1,\dots, \kappa$, $j=1, \dots, n$, is $\kappa$.
The linear constraint (\ref{glin}) is integrable or exact if 
\begin{equation}
	\label{gint}
	g_\nu ({\bf q}, {\dot {\bf q}},t)=\dfrac{d}{dt}f_\nu ({\bf q},t)=
	\sum\limits_{i=1}^n \dfrac{\partial f_\nu}{\partial q_i}{\dot q}_i+\dfrac{\partial f_\nu}{\partial t}
\end{equation}
for a suitable function $f_\nu ({\bf q},t)$.
Holonomic systems are mechanical systems constrained by integrable functions, that is (\ref{gint}) is valid for each $\nu=1, \dots, \kappa$.

\noindent
We also recall that the regularity assumption (\ref{gind}) allow us to write the constraint conditions (\ref{vincg}) in the explicit form: assuming that (except for re--enumerate the variables)
$$
det\,\left(\begin{array}{ccc}
	\dfrac{\partial g_1}{\partial {\dot q}_{n-\kappa+1}}& \dots & 	\dfrac{\partial g_1}{\partial {\dot q}_n}\\
	\vdots & \vdots & \vdots \\
	\dfrac{\partial g_\kappa}{\partial {\dot q}_{n-\kappa+1}}& \dots & 	\dfrac{\partial g_\kappa}{\partial {\dot q}_n}
\end{array}
\right)\not =0
$$
then it is possible to write (\ref{vincg}) as
\begin{equation}
	\label{constrexpl}
	g_\nu={\dot q}_{m+\nu}-\alpha_\nu(q_1, \dots, q_n, {\dot q}_1, \dots, {\dot q}_m, t)=0\qquad \nu=1, \dots, \kappa
\end{equation}
where the generalized velocities $({\dot q}_1,\dots, {\dot q}_m)$, $m=n-\kappa$,  play the role of independent kinetic variables. Correspondingly the virtual displacements $\delta q_1$, $\dots$, $\delta q_m$ can be considered as independent. 

\subsection{Transpositional rule for nonholonomic systems}

\subsubsection{First variation, ${\check{\rm C}}$etaev condition}

\noindent
We recall two different kinds of calculating the variation of a differentiable function  $F({\bf q}, {\dot {\bf q}},t)$:
\begin{equation}
	\label{dcdef}
	\delta^{(c)}F:=\sum\limits_{i=1}^n \dfrac{\partial F}{\partial {\dot q}_i}\delta q_i.
\end{equation}
and
\begin{equation}
	\label{dvdef}
	\delta^{(v)} F:=
	\sum\limits_{i=1}^n \dfrac{\partial F}{\partial q_i}\delta q_i+
	\sum\limits_{i=1}^n \dfrac{\partial F}{\partial {\dot q}_i}\delta {\dot q}_i.
\end{equation}

\noindent
The superscript $(v)$ stands for ``variation'' and recalls the standard method of variational mathematics: the rule comes from considering the Taylor polinomial of $F({\bf q}, {\dot {\bf q}},t)$ up to the first order with respect to the $2n$ variables $({\bf q}, {\dot {\bf q}})=(q_1, \dots, q_n),{\dot {\bf q}}=({\dot q}_1, \dots, {\dot q}_n)$ varied by $(\delta {\bf q},\delta {\dot {\bf q}})$ at a frozen time $t$:
\begin{equation}
	\label{taylor}
	F({\bf q}+\delta {\bf q}, {\dot {\bf q}}+\delta {\dot {\bf q}},t)=F({\bf q}, {\dot {\bf q}},t)+\left( \begin{array}{c} \delta {\bf q}\\\delta {\dot {\bf q}}\end{array} \right)\cdot 
	\left( 
	\begin{array}{c} \nabla_{\bf q}F \\ \nabla_{\dot {\bf q}} F
	\end{array} 
	\right).
\end{equation}
Thus, defining $\delta^{(v)}F=F({\bf q}+\delta {\bf q}, {\dot {\bf q}}+\delta {\dot {\bf q}},t)-F({\bf q}, {\dot {\bf q}},t)$ one gets exactly (\ref{dvdef}). We refer to (\ref{dvdef}) as the first variation of the function $F({\bf q}, {\dot {\bf q}},t)$ at a fixed time $t$.

\noindent
When (\ref{dvdef}) is applied to the constraints (\ref{vincg}), the condition
\begin{equation}
	\label{dv}
	\delta^{(v)}g_\nu =0	
\end{equation}
is usually required in order to deduce the equations of motion from a time--integral variational principle.
	
\noindent
Regarding (\ref{dcdef}), many authors agree on assuming that the constraint (\ref{vincg}) is ideal or perfect if it satisfies the condition  	
\begin{equation}
		\label{dc}
		\delta^{(c)} g_\nu =0
\end{equation}
which is known as ${\check {\rm C}}$etaev condition (the superscript $(c)$ stands for it).

\noindent
Despite the lack of a theoretical justification (in the sense that a way to derive (\ref{dc}) directly from (\ref{vincg}) is lacking, to our knowledge), the skill of the $\check{\rm C}$etaev condition in generalizing the theory of holonomic systems to that of nonholonomic systems is undeniable. Actually, taking into account that $\dfrac{\partial g_\nu}{\partial {\dot q}_i}=\dfrac{\partial f_\nu}{\partial q_i}$ for $g_\nu$ as in (\ref{gint}) (corresponding to the holonomic case), the ${\check{\rm C}}$etaev assumption (\ref{dc}) applied to (\ref{gint}) becomes
	\begin{equation}
		\label{dce}
		\delta^{(c)}g_\nu=\delta^{(c)}_\nu\left(
		\sum\limits_{i=1}^n \dfrac{\partial f_\nu}{\partial q_i}{\dot q}_i+\dfrac{\partial f_\nu}{\partial t}\right) = \sum\limits_{i=1}^n \dfrac{\partial f_\nu}{\partial q_i}\delta q_i=0
	\end{equation}
which corresponds to the standard assumption of ideal geometric constraint for the primitive function $f_\nu$.
More generally, if we consider the linear kinematic constraint (\ref{glin}), the ${\check {\rm C}}$etaev condition (\ref{dc}) writes
	\begin{equation}
		\label{dclin}
		\delta^{(c)}\left(\sum\limits_{j=1}^n a_{\nu,j}({\bf q,t}){\dot q}_j+b_\nu({\bf q},t)\right)
		=	\sum\limits_{j=1}^n a_{\nu,j}\delta q_j=0
	\end{equation}
	which reproduces once again the standard way of defining virtual displacements in the theory of linear kinematic constraints \cite{neimark}.
	
\subsubsection{The transpositional rule}	
	
\noindent
A significant formula links the variations (\ref{dcdef}) and (\ref{dvdef}) to other operations: 
\begin{equation}
	\label{transprule}
	\delta^{(v)} F -\dfrac{d}{dt}\left(\delta^{(c)}F\right)=
	\sum\limits_{i=1}^n 
	\dfrac{\partial F}{\partial {\dot q}_i}\left(\delta {\dot q}_i - \dfrac{d}{dt}\delta q_i\right)-\sum\limits_{i=1}^n {\cal D}_i F \delta q_i
\end{equation}
where $D_i$ is the $i$--th Lagrangian derivative 
\begin{equation}
	\label{derlagr}
	{\cal D}_i F=\dfrac{d}{dt}\dfrac{\partial F}{\partial {\dot q}_i} -\dfrac{\partial F}{\partial q_i}, \qquad i=1, \dots, n.
\end{equation}

\noindent
The identity (\ref{transprule}), known as ``transpositional rule'', is easily inferred by writing (\ref{dvdef}) in the form (via adding and subtracting equal terms)
$$
\delta^{(v)}F=
\dfrac{d}{dt}\left(\sum\limits_{i=1}^n \dfrac{\partial F}{\partial {\dot q}_i}\delta q_i\right)
-\sum\limits_{i=1}^n\left(
\dfrac{d}{dt}\dfrac{\partial F}{\partial {\dot q}_i} -\dfrac{\partial F}{\partial q_i}
\right)\delta q_i
+\sum\limits_{i=1}^n  \dfrac{\partial F}{\partial {\dot q}_i}
\left(\delta {\dot q}_i - \dfrac{d}{dt}\delta q_i\right).
$$
The terms 
\begin{equation}
	\label{tr}
	\delta {\dot q}_i-\dfrac{d}{dt}\left(\delta q_i\right)=\delta \left(\dfrac{d}{dt}q_i\right) - \dfrac{d}{dt}\left(\delta q_i\right), \qquad i=1, \dots, n
\end{equation}
are necessary because of the uncertainty about the commutation of the operators $\delta$ and $\frac{d}{dt}$, which may hold for all coordinates or for part of them. The assumptions
\begin{equation}
	\label{cr}
	\delta {\dot q}_i=\dfrac{d}{dt}(\delta q_i), \qquad i=1, \dots, n
\end{equation}
are the so called commutation relations.

\noindent
The implementation of (\ref{transprule}) to the constraint functions $g_\nu$ of (\ref{vincg}), namely 
\begin{equation}
	\label{transpruleg}
	\delta^{(v)} g_\nu -\dfrac{d}{dt}\left(\delta^{(c)}g_\nu \right)=
	\sum\limits_{i=1}^n 
	\dfrac{\partial g_\nu}{\partial {\dot q}_i}\left(\delta {\dot q}_i - \dfrac{d}{dt}\delta q_i\right)-\sum\limits_{i=1}^n {\cal D}_i g_\nu \delta q_i, \qquad \nu=1, \dots, \kappa
\end{equation}
certainly generates interest, due to the matching of assumptions (\ref{dc}), (\ref{dv}) with the lagrangian derivatives of the constraint functions and of the expressions (\ref{tr}): the vanishing (or not) of these two latter quantities strongly depend on the type of nonholonomic system we are considering.

\subsubsection{Transpositional rule for constraints in explicit form}

\noindent
From (\ref{constrexpl})--(\ref{dvdef}) and (\ref{derlagr}) one easily gets
\begin{equation}
	\label{formexpl}
%	\left.
\begin{array}{l}
	\delta^{(c)}g_\nu 
	= \delta^{(c)}({\dot q}_{m+\nu}-\alpha_\nu)
%	=\sum\limits_{i=1}^n \left( \dfrac{\partial {\dot q}_{m+\nu}}{\partial {\dot q}_i}\delta q_i
%	- \dfrac{\partial \alpha_\nu}{\partial {\dot q}_i}\right)\delta q_i
	=\delta q_{m+\nu}
	-\sum\limits_{r=1}^m
	\dfrac{\partial \alpha_\nu}{\partial {\dot q}_r}\delta q_r\\
	\\
	\delta^{(v)}g_\nu = \delta^{(v)}({\dot q}_{m+\nu}-\alpha_\nu)=-\sum\limits_{i=1}^n \dfrac{\partial \alpha_\nu}{\partial q_i}\delta q_i
	+\delta {\dot q}_{m+\nu}
	-\sum\limits_{r=1}^m\dfrac{\partial \alpha_\nu}{\partial {\dot q}_r}\delta {\dot q}_r
	\\
	\\
{\cal D}_i g_\nu=\left\{ \begin{array}{ll}-{\cal D}_i\alpha_\nu
	 &for\;i=1, \dots, m\\
	\\
	\dfrac{\partial \alpha_\nu}{\partial q_i}&for\;i=m+1, \dots, n
	\end{array}\right.
%	=\sum\limits_{i=1}^n 
%	\left( \dfrac{\partial}{\partial q_i}({\dot q}_{m+\nu}-\alpha_\nu)\delta q_i+
%	\dfrac{\partial}{\partial {\dot q}_i}({\dot q}_{m+\nu}-\alpha_\nu)\delta {\dot q}_i\right)
\end{array}
%\right\}
\end{equation}
wherefrom the transpositional rule in the case of the explicit expressions (\ref{constrexpl}) is deduced:
\begin{eqnarray}
	\nonumber
	\delta^{(v)} g_\nu -\dfrac{d}{dt}\left(\delta^{(c)}g_\nu\right)&=&
	\delta {\dot q}_{m+\nu}-\dfrac{d}{dt}\left( \delta q_{m+\nu}\right)
	-\sum\limits_{r=1}^m \dfrac{\partial \alpha_\nu}{\partial {\dot q}_r}\left(\delta {\dot q}_r - \dfrac{d}{dt}\delta q_r\right)\\
	\label{transpruleexpl}
	&+&\sum\limits_{r=1}^m  {\cal D}_r \alpha_\nu \delta q_r 
	-\sum\limits_{\mu=1}^\kappa
	\dfrac{\partial \alpha_\nu}{\partial q_{m+\nu}}\delta q_{m+\mu}, \qquad \nu=1, \dots, \kappa.
\end{eqnarray}
The same formulae (\ref{formexpl}) allow to assert that $\delta^{(c)}g_\nu =0$ (condition (\ref{dc})) is equivalent to 
\begin{equation}
	\label{dcdip}
	\delta q_{m+\nu}=\sum\limits_{r=1}^m \dfrac{\partial \alpha_\nu}{\partial {\dot q}_r}\delta q_r, \qquad \nu=1,\dots, \kappa
\end{equation}
and $\delta^{(v)}g_\nu =0$ (condition (\ref{dv})) is equivalent to
\begin{equation}
	\label{dvdip}
	\delta {\dot q}_{m+\nu}=
	\sum\limits_{i=1}^n \dfrac{\partial \alpha_\nu}{\partial q_i}\delta q_i+
	\sum\limits_{r=1}^m \dfrac{\partial \alpha_\nu}{\partial {\dot q}_r}\delta {\dot q}_r, \qquad \nu=1,\dots, \kappa.
\end{equation}

\subsection{Equations of motion for nonholonomic systems (a hint)}

\noindent
Although we will not deal with the question of formulating the equations of motion for systems constrained by the conditions (\ref{vincg}), 
it is at any rate suitable to mention some key points, in order to better understand the role of the conditions (\ref{dc}), (\ref{dv}) and of the formula (\ref{transprule}). We identify two relevant approaches (even though alternative methods are possible): the d'Alembert--Lagrange principle and the generalization of the Hamilton principle.

\subsubsection{Equations of motion via the d'Alembert--Lagrange principle}

\noindent
By the d'Alembert principle one assumes that (see for instance \cite{neimark})
	\begin{equation}%pag. 91 NF
		\label{dal}
		\sum\limits_{i=1}^n \left(
		\dfrac{d}{dt}\dfrac{\partial {\cal L}}{\partial {\dot q}_i}-\dfrac{\partial {\cal L}}{\partial q_i}\right) \delta q_i=0 
	\end{equation}
holds for any set of virtual displacements $(\delta q_1, \dots, \delta q_n)$ of the system, that is any variation of the coordinates which is compatible with the constraints (\ref{vincg}) at a given instant of time.
The terms of the matter move to the question of identifying the proper class of virtual displacements for the kinematic constraints (\ref{vincg}), so that a differential form of equations of motion can be deduced from (\ref{dal}).
A straightforward way is based on the ${\check {\rm C}}$etaev condition (\ref{dc}):
since the constraint functions $g_1$, $\dots$, $g_\nu$ are given, (\ref{dc}) is used to identify the whole class of virtual displacements $\delta {\bf q}$ as the $(n-\kappa)$--dimensional vector space ${\cal W}^T$ orthogonal to the $\kappa$--dimensional space ${\cal W}=span (\frac{\partial g_1}{\partial {\dot {\bf q}}}, \dots, \frac{\partial g_\kappa}{\partial {\dot {\bf q}}})$ (the $\kappa$ vectors are independent by virtue of (\ref{gind})).
Owing to arbitrariness of ${\delta {\bf q}}$ in ${\cal W}^T$, from (\ref{dal}) we deduce that 
$\frac{d}{dt}\frac{\partial {\cal L}}{\partial {\dot {\bf q}}}-\frac{\partial {\cal L}}{\partial {\bf q}}$ belongs to ${\cal W}$, namely of the type $\sum\limits_{\nu=1}^\kappa \mu_\nu \frac{\partial g_\nu}{\partial {\dot {\bf q}}}$ for some unknown multipliers $\mu_\nu$, thereby  the equations of motion take the form
\begin{equation}
	\label{eqmotodc}
	\dfrac{d}{dt}\dfrac{\partial {\cal L}}{\partial {\dot q}_i}-\dfrac{\partial {\cal L}}{\partial q_i}=\sum\limits_{\nu=1}^\kappa \mu_\nu 
	\dfrac{\partial g_\nu}{\partial {\dot q}_i}, \qquad i=1, \dots, n
\end{equation}
and, joined to the constraint equations (\ref{vincg}), they form a system of $n+\kappa$ equations in the $n+\kappa$ unknown functions ${\bf q}(t)$, $\mu_1(t)$, $\dots$, $\mu_\nu(t)$.

\noindent
Notice that neither condition (\ref{dc}) nor an assumption about the differences (\ref{tr}) are necessary in order to infer the equations of motion (\ref{eqmotodc}).

\subsubsection{The equations of motions via a time--integral variational principle}

\noindent
The generalization to nonholonomic systems of the Hamilton's principle, which leads for holonomic systems to the Lagrange's equations of motion starting from the stationarity of the action functional, is a problematic issue and an unsolved question, as far as we know.
We refer to \cite{cronstrom},  \cite{flanneryenigma}, \cite{krup4}, \cite{lemos}, \cite{rumy2000} for interesting discussions and results and we select some essential points in view of our investigation.

\noindent
In extending the Hamilton principle to nonholonomic systems, two aspects are taken into account:

\begin{itemize}
\item{} the stationarity $\delta \int\limits_{t_0}^{t_1}{\cal L}dt=0$ is replaced, owing to the restrictions (\ref{vincg}), by 
\begin{equation}
	\label{hsp}
	\delta	\bigintsss_{t_0}^{t_1} \left({\cal L}({\bf q}, {\dot {\bf q}},t)+\sum\limits_{\nu=1}^\kappa \lambda_\nu(t) g_\nu({\bf q}, {\dot {\bf q}},t) \right)dt=0
\end{equation}
with $\lambda_1$, $\dots$, $\lambda_\kappa$ Lagrange multipliers.
\item{} the possible non--validity of the commutation relations (\ref{cr}) leads to a new version of the Hamilton's principle $\delta \int\limits_{t_1}^{t_2}Ldt=0$ (where $L$ denotes a generic Lagrangian function): in \cite{neimark} the principle of stationary action is presented in the more general form
	\begin{equation}
		\label{sa}
		\int\limits_{t_0}^{t_1} \left(
		\delta L - \sum\limits_{i=1}^n \dfrac{\partial L}{\partial {\dot q}_i}\left(\delta {\dot q}_i- \dfrac{d}{dt}\left(\delta q_i\right)\right)\right)dt=0
	\end{equation}
	where $\delta L=\sum\limits_{i=1}^n \frac{\partial L}{\partial q_i}\delta q_i+
	\sum\limits_{i=1}^n\frac{\partial L}{\partial {\dot q}_i}\delta {\dot q}_i$
	is calculated in accordance with (\ref{dv}). 
	The point of view adopted for the differences $\delta {\dot q}_i - \frac{d}{dt}(\delta q_i)$ does affect the special form of the principle.
\end{itemize}

\begin{rem}
We date back the principle (\ref{hsp}) to \cite{ray},  which gives rise to the so called vakonomic mechanics \cite{arnoldkozlov}.	

\noindent
The generalization (\ref{sa}) of the Hamilton's principle (Hamilton--Suslov principle, \cite{neimark}, \cite{llibre}) simply originates from the integration w.~.r.~t.~time of (\ref{dal}) in the interval $[t_0, t_1]$ and from the identity
$$
\sum\limits_{i=1}^n\dfrac{d}{dt}\left(\dfrac{\partial L}{\partial {\dot q}_i}\right)\delta q_i=
\sum\limits_{i=1}^n \dfrac{d}{dt}\left(\dfrac{\partial L}{\partial {\dot q}_i}\delta q_i\right)
-\sum\limits_{i=1}^n\dfrac{\partial L}{\partial {\dot q}_i}\delta {\dot q}_i  -\sum\limits_{i=1}^n \dfrac{\partial L}{\partial {\dot q}_i}\left(\dfrac{d}{dt}\left(\delta q_j\right)-\delta {\dot q}_i\right)
$$
(it is assumed that the variations $\delta q_i$ vanish at the ends of the interval of integration).
\end{rem}

\noindent
The combination of (\ref{hsp}) and (\ref{sa}), namely the introduction of the constrained Lagrangian function 
${\cal L}+\sum\limits_{\nu=1}^\kappa \lambda_\nu g_\nu$ in the functional (\ref{sa}) leads to 
\begin{equation}
	\label{hs}
	\int\limits_{t_0}^{t_1} \left(
	\delta \left({\cal L}+\sum\limits_{\nu=1}^\kappa \lambda_\nu g_\nu\right)
	-\sum\limits_{i=1}^n \dfrac{\partial}{\partial {\dot q}_i}
	 \left({\cal L}+\sum\limits_{\nu=1}^\kappa \lambda_\nu g_\nu\right)
	\left(\delta {\dot q}_i-\dfrac{d}{dt}\left(\delta q_i\right)\right)\right)dt=0
\end{equation}
If one assumes that the transpositional relations (\ref{tr}) can be given in the form \cite{llibre}
\begin{equation}
	\label{tra}
\delta {\dot q}_i-\dfrac{d}{dt}\delta q_i	= \sum\limits_{j=1}^n W_{i,j}({\bf q}, {\dot {\bf q}}, \bfq2dot^{..}, t)\delta q_j, \qquad i=1, \dots, n
\end{equation}
then the extended Hamiltonian principle (\ref{hs}) leads to the equations of motion \cite{llibre}
\begin{equation}
	\label{eqllibre}
	{\cal D}_i{\cal L}_R -  \sum\limits_{j=1}^n W_{j,i}\dfrac{\partial {\cal L}_R}{\partial {\dot q}_j}=0, \qquad i=1, \dots, n
\end{equation}	
where we shortened notations by defining the constrained Lagrangian  
${\cal L}_R={\cal L}+\sum\limits_{\nu=1}^\kappa \lambda_\nu g_\nu$.
Equations (\ref{eqllibre}) have to be joined to the constraint equations (\ref{vincg}), in order to form a system of $n+\kappa$ equations in the $n+\kappa$ unknowns given by $q_i(t)$ , $i=1, \dots, n$, and $\lambda_\nu(t)$, $\nu=1, \dots, \kappa$. 

\begin{rem}
	If $W_{i,j}=0$ for each $i,j=1, \dots, n$, then (\ref{tra}) are the commutation relations (\ref{cr}): in this case the equations of motion (\ref{eqllibre}) reduce to ${\cal D}_i {\cal L}_R=0$ and they match with the ones proposed in \cite{ray}, namely
\begin{equation}
		\label{eqray}
		\dfrac{d}{dt}\dfrac{\partial {\cal L}}{\partial {\dot q}_i}-\dfrac{\partial {\cal L}}{\partial q_i}=\sum\limits_{\nu=1}^\kappa \left(\lambda_\nu \left(
		\dfrac{\partial g_\nu}{\partial q_i}-\dfrac{d}{dt}\dfrac{\partial g_\nu}{\partial {\dot q}_i}\right)-{\dot \lambda}_\nu \dfrac{\partial g_\nu}{\partial {\dot q}_i}\right), \qquad i=1, \dots, n
	\end{equation}
which differ from (\ref{eqmotodc}). It is evident that if the constraints (\ref{vincg}) are absent, then in both cases (\ref{eqmotodc}) and (\ref{eqray}) the equations of motion reduce to the standard Lagrange equations $\frac{d}{dt}\frac{\partial {\cal L}}{\partial {\dot q}_i}-\frac{\partial {\cal L}}{\partial q_i}=0$, $i=1, \dots, n$.
\end{rem}	
	
\noindent
The generalization (\ref{eqllibre}) of (\ref{eqray}), known as modificated vakonomic mechanics \cite{llibre}, \cite{pastore}, aims to encompass nonholonomic systems in the framework of the vakonomic mechanics, which at the primary stage (\ref{eqray}) is not applicable to them, according to a certain number of papers (see for instance \cite{lemos}, \cite{liromp}).
The question arising with the hypothesis (\ref{tra}) concerns the determination of the coefficients $W_{i,j}$: in \cite{pastore} an extensive discussion deals with this theme.
The equalities (\ref{tra}) are in relation to the transpositional rule, which will be presented shortly after and where our analysis will focus on.

\noindent
An operative assumption introduced in \cite{llibre} consists in hypothesizing the relations
\begin{equation}
\label{llibre2}
{\cal D}_i g_\nu - \sum\limits_{j=1}^n W_{j,i}\dfrac{\partial g_\nu}{\partial {\dot q}_j}=0, \qquad i=1, \dots, n, \quad \nu=1, \dots, \kappa.
\end{equation}
In this case, the equations of motion (\ref{eqllibre}) reduce to
\begin{equation}
	\label{eqllibre2}
	{\cal D}_i{\cal L} - \sum\limits_{j=1}^n W_{j,i}\dfrac{\partial {\cal L}}{\partial {\dot q}_j}=-\sum\limits_{j=1}^n \sum\limits_{\nu=1}^\kappa W_{j,i}{\dot \lambda}_\nu \dfrac{\partial g_\nu}{\partial {\dot q}_j}, \qquad i=1, \dots, n.
\end{equation}	
We will comment such possibility further on.

\section{The mathematical aspect}

\noindent
For the sake of clarity, we gather assumptions (\ref{dc}), (\ref{dv}), (\ref{cr}) and \ref{tra}) in the following list:
\begin{equation}
	\label{abc0}
\begin{array}{llll}
\delta^{(c)} g_\nu =
\sum\limits_{i=1}^n \dfrac{\partial g_\nu}{\partial {\dot q}_i}\delta q_i=0, 
& \nu=1, \dots, \kappa & &\qquad \quad (A)\\
\\
\delta^{(v)} g_\nu=
\sum\limits_{i=1}^n \dfrac{\partial g_\nu}{\partial q_i}\delta q_i+
\sum\limits_{i=1}^n \dfrac{\partial g_\nu}{\partial {\dot q}_i}\delta {\dot q}_i=0,  
&\nu=1, \dots, \kappa & &\qquad \quad (B)\\
\\
\delta {\dot q}_i- \dfrac{d}{dt}\left(\delta q_i\right)=0,  &i=1, \dots, n & & \qquad \quad (C_0)\\
\\
\delta {\dot q}_i-\dfrac{d}{dt}\left(\delta q_i\right)=\sum\limits_{j=1}^n W_{i,j}\delta q_j, &i=1, \dots, n & & \qquad \quad (C)
\end{array}
\end{equation}
With respect to (\ref{transpruleexpl}), we add a halfway point of view between $(C_0)$ and $(C)$:
\begin{equation}
	\label{c1}
	\delta {\dot q}_r- \dfrac{d}{dt}\left(\delta q_r\right)=0,  \qquad \qquad \qquad \quad r=1, \dots, m \qquad \qquad \qquad  (C_1)
\end{equation}
asserting that the commutation relations (\ref{cr}) hold only for the independent variables $(q_1, \dots, q_m)$.
The standpoint in front of the previous conditions defines the specific approach used to treat nonholonomic systems. 

\noindent
In Paragraph 1.3.1 we outlined that the ${\check {\rm C}}$etaev assumption alone (that is without a pronouncement about the variations $\delta^{(v)}g_\nu$ and the commutations (\ref{cr})) allows to formulate the equations of motion in the form (\ref{eqmotodc}). This case corresponds to assume $(A)$ ignoring $(B)$, $(C)$ and the rule (\ref{tr}) is 
	\begin{equation}
		\label{transprulea}	
		\delta^{(v)}g_\nu=	\sum\limits_{i=1}^n 
		\dfrac{\partial g_\nu}{\partial {\dot q}_i}\left(\delta {\dot q}_i - \dfrac{d}{dt}\delta q_i\right)-\sum\limits_{i=1}^n {\cal D}_i g_\nu \delta q_i
	\end{equation}
	which can be used to calculate the variations $\delta^{(v)}g_\nu$, generally not null in default of further assumptions. If it is known that (\ref{derlagrsomma0}) holds,then (\ref{transprulea}) reduces to
	\begin{equation}
		\label{transprulead0}	
		\delta^{(v)}g_\nu=	\sum\limits_{i=1}^n 
		\dfrac{\partial g_\nu}{\partial {\dot q}_i}\left(\delta {\dot q}_i - \dfrac{d}{dt}\delta q_i\right).
	\end{equation}

\noindent
	The combination of $(A)$ and $(C_0)$ corresponds to the H\"older principle condition (see \cite{rumycet}), whereas the coupling of $(B)$ and $(C_0)$ is at the base of the vakonomics equations of motion (\ref{eqray}) (see \cite{arnoldkozlov}). The broader assumption $(C)$ joined with $(B)$ produces the recent approaches of generalizing the vakonomic mechanics.

\subsection{Necessary conditions for commutation}

\noindent
We start by some evident consequences (\ref{transpruleg}), which points out in each case $(A)$,  $(B)$ or the combination of them, which we refer to as assumption $(AB)$, the necessary conditions for the commutation relations (\ref{cr}).

\begin{prop}
Assume that the commutation relations $(C_0)$ hold. Then:
\begin{eqnarray}
	\label{dvderlagr}
	\delta^{(v)}g_\nu=-\sum\limits_{i=1}^n {\cal D}_ig_\nu \delta q_i
&\textrm{in the case}\;(A),
\\
\label{dcderlagr}
\dfrac{d}{dt}\left(\delta^{(c)}g_\nu\right) =\sum\limits_{i=1}^n {\cal D}_i g_\nu \delta q_i
&\textrm{in the case}\;(B),
\\
\label{derlagr0}
\sum\limits_{i=1}^n {\cal D}_i g_\nu \delta q_i=0
&\textrm{in the case}\;(AB),
\end{eqnarray}
\end{prop}
	
\noindent
Each of the three identities (\ref{dcderlagr})--(\ref{derlagr0}) is a necessary condition for the commutation $\delta {\dot q}_j-\frac{d}{dt}\delta q_j$, $j=1, \dots, n$, in their respective cases
in their respective cases $(A)$ or $(B)$ or $(AB)$.
Notice that the reverse sense of the Proposition is not true: if either (\ref{dvderlagr}) or
(\ref{dcderlagr}) or (\ref{derlagr0}) holds  then 
$\sum\limits_{i=1}^n 
\frac{\partial g_\nu}{\partial {\dot q}_i}(\delta {\dot q}_i-\frac{d}{dt}\delta q_i)=0$, but this does not imply 
$(C_0)$, even though the latter equality holds for any $\nu=1, \dots, \kappa$ (we recall that $\kappa<n$). 

\noindent
We remark that, whenever $(C_0)$ is assumed, the condition
\begin{equation}
\label{derlagrsomma0}
\sum\limits_{i=1}^n {\cal D}_ig_\nu \delta q_i=0
\end{equation}
makes us write (\ref{transpruleg}) as
$$	
\delta^{(v)} g_\nu =\dfrac{d}{dt}\left(\delta^{(c)}g_\nu \right)
$$
so that in the case $(A)$ (namely $\delta^{(c)}g_\nu=0$) also $\delta^{(v)}g_\nu=0$. On the other hand, in the case $(B)$ the previous relation gives $\delta^{(c)}g_\nu=constant$ whose value can be deduced by choosing the null displacements $\delta q_j=0$, getting $constant=0$.
We summarize the argument by the following

\begin{prop}
Assume that $(C_0)$ holds. Then the identity (\ref{derlagrsomma0}) makes the two conditions $(A)$ and $(B)$ simultaneously zero or not zero.
\end{prop}

\noindent
We see that  condition (\ref{derlagrsomma0}) offers a special case: further on we will investigate the identification of the class of functions which verify such a condition.

\subsection{The double assumption $(A)$ and $(B)$}

\noindent
The assumption $(A)$ plays, to say, a physical role and intends to identify ideal constraints by means of virtual displacements. On the other hand, the assumption $(B)$ is mainly related to the implementation of variational principles and to the mathematical techniques of calculus of variation.
The combination of $(A)$ and $(B)$ (which we continue to denote by $(AB)$) is largely considered in literature either in classical treatises \cite{pars} \cite{lurie} and in models proposed in literature.
A diffuse skepticism concerns the coupling $(AB)$ if this occurs in the frame of the commutation relations $(C_0)$: in this case only integralbe contraints can be considered. According to some new approaches \cite{llibre} \cite{pastore}, the replacement of $(C_0)$ with a more general assumption $(C)$ (transpositional rule (\ref{tra})),  leads to a coexistence of $(AB)$ and $(C)$ even in a context of nonholonomic systems. 

\noindent
In simple terms, we are exploring the opposite sense of Corollary 1: assuming $(AB)$, what can be said about the two properties $(C_0)$ and (\ref{derlagrsomma0})?

\noindent
An evidence of the case we are studying is the following

\begin{propr}
Assume that both $(AB)$ holds. Then if the commutation relations $(C_0)$ are verified, then condition (\ref{derlagrsomma0}) must hold for any $\nu=1, \dots, \kappa$.
\end{propr}
Indeed, if $(A)$ and $(B)$ are both in force then the transpositional rule (\ref{transpruleg}) reduces to 
\begin{equation}
	\label{transpruleab}	
	\sum\limits_{i=1}^n 
	\dfrac{\partial g_\nu}{\partial {\dot q}_i}\left(\delta {\dot q}_i - \dfrac{d}{dt}\delta q_i\right)=\sum\limits_{i=1}^n {\cal D}_i g_\nu \delta q_i
\end{equation}
and the sum on the left side is zero if $(C_0)$ is verified, whence (\ref{derlagrsomma0}) must hold.

\noindent
We see that the identity (\ref{derlagrsomma0}) is a necessary condition for the commutation relation (\ref{cr}).
As we already remarked just after Proposition 1, the same identity does not imply the commutation $(C_0)$, but only 
\begin{equation}
	\label{abc1}
\sum\limits_{i=1}^n 
\dfrac{\partial g_\nu}{\partial {\dot q}_i}\left(\delta {\dot q}_i 
- \dfrac{d}{dt}\delta q_i\right)=0.
\end{equation}

\noindent
A characterization of the identity (\ref{derlagrsomma0}) can be given by only assuming $(A)$:

\begin{prop}
	Assume that $\delta^{(c)}g_\nu=0$ for each $\nu=1, \dots, \kappa$. Then the equality (\ref{derlagrsomma0}) holds if and only if 
	\begin{equation}
		\label{zero2}{\cal D}_i g_\nu =\sum\limits_{\mu=1}^\kappa\varrho_\mu^{(\nu)}
		\dfrac{\partial g_\mu}{\partial {\dot q}_i}\qquad i=1, \dots, n
	\end{equation}
	for some real coefficients $\varrho_\mu^{(\nu)}({\bf q}, {\dot {\bf q}},t)$, $\mu=1, \dots, \kappa$, $i=1, \dots, n$.
\end{prop}

\noindent
{\bf Proof}. 
If $\sum\limits_{i=1}^n {\cal D}_i g_\nu \delta q_i=0$ for any displacement verifying $\sum\limits_{i=1}^n \frac{\partial q_\nu}{\partial {\dot q}_i}\delta q_i=0$, then at each state $({\bf q}, {\dot {\bf q}},t)$ 
the linear dependence (\ref{zero2}) on the vectors $\frac{\partial g_\nu}{\partial {\dot {\bf q}}}$ must be true, owing to (\ref{gind}). 
Conversely, if (\ref{zero2}) holds then, multiplying it by $\delta q_i$ and summing with respect to $i=1, \dots, n$, one finds
$$
\sum\limits_{i=1}^n {\cal D}_i g_\nu \delta q_i=\sum\limits_{\mu=1}^\kappa\varrho_\mu^{(\nu)}
\sum\limits_{i=1}^n \dfrac{\partial g_\mu}{\partial {\dot q}_i}\delta q_i=0
$$
by virtue of condition $(A)$.
$\quad \square$

\begin{rem}
By virtue of Properties 2 and 3, the identity (\ref{zero2}) is a necessary condition for the commutation $(C_0)$. This leads to think that only linear kinematic constraints (i.~e.~of the type (\ref{glin})) potentially access the commutation relations (\ref{cr}), since otherwise the left side of (\ref{zero2}) would contain the second derivatives ${\q2dot^{..}}_i$, which are absent in the right side.
\end{rem}

\subsubsection{The set of displacements $\delta {\bf q}$ and $\delta {\dot {\bf q}}$}

\noindent
The main results are arranged in Propositions 1 and 2 and aim at placing the dispacements $\delta {\bf q}$ and ${\delta {\dot \bf q}}$ in specific linear spaces, whenever $(AB)$ holds.

\begin{prop}
At any given state $({\bf q}, {\dot {\bf q}},t)$, the set of displacements $\delta {\bf q}$ consistent with $(AB)$ coincides with the $(n-\kappa)$--dimensional vector space ${\Bbb V}^{(n-\kappa)}$ orthogonal to that generated by the vectors $\dfrac{\partial g_\nu}{\partial {\dot {\bf q}}}$, $\nu=1, \dots, \kappa$. The set of variations $\delta {\dot {\bf q}}$ consistent with $(AB)$ is an affine space obtained by a traslation of ${\Bbb V}^{(n-\kappa)}$.
\end{prop}

\noindent
{\bf Proof}. It is convenient to switch to a vector--matrix notation, by defining 
the two $\kappa\times n$ jacobian matrices
\begin{equation}
	\label{matrab}
A({\bf q}, {\dot {\bf q}},t)=\left.\left(\dfrac{\partial g_\nu}{\partial q_j}\right)\right\vert_{\begin{array}{l}\nu=1, \dots, \kappa\\j=1, \dots, n\end{array}} \qquad 
B({\bf q}, {\dot {\bf q}},t)=\left.\left(\dfrac{\partial g_\nu}{\partial {\dot q}_j}\right)\right\vert_{\begin{array}{l}\nu=1, \dots, \kappa\\j=1, \dots, n\end{array}}
\end{equation}
and the two $n\times 1$ column vectors 
$$
\delta {\bf q}=(\delta q_1, \dots, \delta q_n)^T, \qquad
\delta {\dot {\bf q}}=(\delta {\dot q}_1, \dots, \delta {\dot q}_n)^T.
$$
In this way, the fulfillment of $(A)$ and $(B)$ corresponds at each state $({\bf q}, {\dot {\bf q}},t)$ to the linear system
\begin{equation}
\label{abvett}
\left\{
\begin{array}{ll}
B\delta {\bf q}={\bf 0}& \qquad \qquad (A)\\
	\\
A\delta {\bf q}+B\delta {\dot {\bf q}}={\bf 0}& \qquad \qquad (B)\\
\end{array}
\right.
\end{equation}
of $2\kappa$ equations for the $2n$ unknown variations (at that state) $(\delta {\bf q}, 
{\delta {\dot {\bf q}}})$. Obviuosly $(C_0)$ is 
\begin{equation}
	\label{c0vett}
\dfrac{d}{dt}(\delta {\bf q})=\delta {\dot {\bf q}}.
\end{equation}
The non--singularity assumption (\ref{gind}) corresponds to $rank\,B=\kappa$ and the first $\kappa$ equations in (\ref{abvett}) imply
\begin{equation}
\label{dq}
\delta {\bf q}=\sum\limits_{r=1}^m \sigma_r({\bf q}, {\dot {\bf q}},t) {\bf w}_r({\bf q}, {\dot {\bf q}},t)
\end{equation}
(we recall that $m=n-\kappa$) where $\sigma_r$, $r=1, \dots, m$, are real--valued functions and ${\bf w}_r$, $r=1, \dots, m$, are $m$ indipendent vectors in ${\Bbb R}^n$ forming a basis of the orthogonal complement to the space generated by the rows of (\ref{gind}), that is
\begin{equation}
\label{w}
\langle {\bf w}_1, \dots, {\bf w}_m\rangle=
\left\langle 
\dfrac{\partial g_1}{\partial {\dot {\bf q}}}, \dots, 
\dfrac{\partial g_\kappa}{\partial {\dot {\bf q}}}\right\rangle^\perp\qquad \textrm{at each state}\;\;({\bf q}, {\dot {\bf q}},t)
\end{equation}
and the first part of the Proposition is proved.
Consider now the second group of $\kappa$ equations in (\ref{abvett}), which take the form, owing to (\ref{dq}),
\begin{equation}
\label{abvett2}
B\delta {\dot {\bf q}}=-A\delta {\bf q}=-\sum\limits_{r=1}^m \sigma_r A {\bf w}_r.
\end{equation}
At each state $({\bf q}, {\dot {\bf q}},t)$ and for a $\delta {\bf q}$ fixed as in (\ref{dq}), equations (\ref{abvett2}) form a linear nonhomogeneous system of $\kappa$ equations in the $n$ unknowns $\delta {\dot {\bf q}}$: the solutions are 
\begin{equation}
\label{deltadotq}
\delta {\dot {\bf q}}=\sum\limits_{s=1}^m \psi_s {\bf w}_s -\sum_{r=1}^m \sigma_r B^+A {\bf w}_r
\end{equation}
where $\psi_s({\bf q}, {\dot {\bf q}},t)$ are arbitrary and the first sum represents all the solutions of the corresponding homogeneous system $B {\bf x}={\bf 0}$, whereas the second sum corresponds to a particular solution of the nonhomogeneous system, written by any $n\times \kappa$ matrix $B^+$ right inverse of $B$, that is $BB^+={\Bbb I}_\kappa$, identity matrix of order $\kappa$ (we recall that $rank B=\kappa$ and existence of $B^+$ are equivalent conditions and that $B^+{\bf b}$ is a particular solution of the system $B{\bf x}={\bf b}$, see \cite{golub}). 
Taking into account that the coefficients $\psi_s$, $s=1, \dots, m$, are arbitrary, 
the expression (\ref{deltadotq}) shows that for a given $\delta {\bf q}$ the totality of $\delta {\dot {\bf q}}\,$s compatible with (\ref{abvett}) is the affine space ${\Bbb A}^{(n-\kappa)}$ achieved by translating the space ${\Bbb V}^{(n-\kappa)}$  This proves the second part of the Proposition. $\quad \square$

\noindent
We now write (\ref{transpruleab}) in vector--matrix notation (see also (\ref{matrab}))
\begin{equation}
\label{transpruleabmatr}
B\left(\delta {\dot {\bf q}}-\dfrac{d}{dt}(\delta {\bf q})\right)=D\delta {\bf q}=({\dot B}-A)\delta {\bf q}
\end{equation}
where $D({\bf q}, {\dot {\bf q}},t)$ is the $\kappa\times n$--matrix with entries 
$D_{\nu,i}={\cal D}_i g_\nu$, $\nu=1, \dots, \kappa$, $i=1, \dots, n$ (${\cal D}_i$ is the Lagrangian derivative defined in (\ref{derlagr})). We gather the main results in the following

\begin{prop}
Assume $(AB)$. Then
	\begin{itemize}
		\item{} $(C_0)$ implies $D \delta {\bf q}={\bf 0}$,
		\item{} $D {\delta {\bf q}}={\bf 0}$ implies 
		\begin{description}
			\item[$(i)$] $\delta {\dot {\bf q}}-\dfrac{d}{dt}(\delta {\bf q})\in  span\,
			\langle {\bf w}_1,\dots,{\bf w}_m\rangle$,
			\item[$(ii)$] $A\delta {\bf q}+B \delta {\dot {\bf q}}=A\delta {\bf q}+B \dfrac{d}{dt}(\delta {\bf q})={\bf 0}$,
			\item[$(iii)$] $\dfrac{d}{dt} (\delta {\bf q})\in {\Bbb A^{(n-\kappa)}}$
		\end{description}
where ${\Bbb A}^{(n-\kappa)}$ is the affine space defined by (\ref{deltadotq}).
	\end{itemize}
\end{prop}

\noindent
{\bf Proof}. The first statement follows immediately from (\ref{c0vett}) and (\ref{transpruleabmatr}).
The same equality shows that if $D\delta {\bf q}={\bf 0}$ then $B(\delta {\dot {\bf q}}-\frac{d}{dt}(\delta {\bf q}))={\bf 0}$ therefore $\delta {\dot {\bf q}}-\frac{d}{dt}\delta {\bf q}$ is a linear combination of the vectors (\ref{w}) (from which point $(i)$), owing to the same argument used for (\ref{dq}). 
Statement $(ii)$ put together condition $(B)$ and $B\delta {\bf q}=B\frac{d}{dt}(\delta {\bf q})$.
In order to check $(iii)$ we calculate the time derivative of condition $(A)$ (see (\ref{abvett})):
$$
\dfrac{d}{dt}(B \delta {\bf q})={\dot B}\delta {\bf q}+B \dfrac{d}{dt}(\delta {\bf q})={\bf 0}.
$$
We compare the latter equality with condition $(B)$ of (\ref{abvett}) written as
$$
\begin{array}{lll}
\textrm{\circled{1}} \quad B\delta {\dot {\bf q}}=-A\delta {\bf q}  &\qquad \qquad &
\textrm{\circled{2}} \quad B \dfrac{d}{dt}{\delta {\bf q}}=-{\dot B}\delta {\bf q}
\end{array}
$$
We look at \circled{1} and \circled{2} as linear relations for $\delta {\dot {\bf q}}$ and $\frac{d}{dt}\delta {\bf q}$ respectively: as seen above, the variations $\delta {\dot {\bf q}}$ satisfying \circled{1} are all and only the elements of the affine space ${\Bbb A}^{(n-\kappa)}$ described by (\ref{deltadotq}). By virtue of the current assumption $D{\bf q}={\bf 0}$ implying $A\delta {\bf q}=-{\dot B}\delta {\bf q}$, the linear relation \circle{2} for $\frac{d}{dt}(\delta {\bf q})$ is identical to \circled{1} (written for $\delta {\bf q}$): therefore, the vector $\dfrac{d}{dt}(\delta {\bf q})$ must be one of the elements of the affine space ${\Bbb A}^{(n-\kappa)}$. $\quad\square$.
	
\noindent
We briefly comment the results of Proposition 2.

\noindent	
As we already remarked, $D\delta {\bf q}={\bf 0}$ (that is (\ref{derlagrsomma0})) is a necessary but not sufficient condition in order that the commutation relation $(C_0)$ is fufilled.
The assumption $D\delta {\bf q} = {\bf 0}$ situates the difference $\delta {\dot {\bf q}}-\frac{d}{dt}(\delta {\bf q})$ in the space orthogonal to the vectors $\frac{\partial g_\nu}{\partial {\dot {\bf q}}}$, $\nu=1, \dots, \kappa$ (point $(i)$ of the Proposition).
Point $(ii)$ means that the variation $\delta^{(v)}g_\nu$ defined in (\ref{dv}) can be calculated either with $\delta {\dot q}_i$ or $\frac{d}{dt}(\delta {\dot q}_i)$, $i=1, \dots, n$.
We also remark that if $D\delta {\bf q}\not = {\bf 0}$, then the the set of the vectors $\delta {\dot {\bf q}}$ verifying \circled{1} (see the proof of Proposition 2) and the set of those verifying \circled{2} are disjointed: actually, the linear relations \circled{1}, \circled{2} show the same matrix $B$ in the left sides, but $-A\delta {\bf q}\not =-{\dot B}\delta {\bf q}$ at the right sides.

\noindent
We finally move our attention to the more general assumption $(C)$ stated in (\ref{tra}), which has been recently considered for improving the techniques of vakonomics mechanics in order to encompass nonholonomic systems. In  order to identify the coefficients $W_{i,j}$ of (\ref{tra}), in \cite{llibre} the relations (\ref{llibre2}) are assumed to hold for each $i=1, \dots, n$ and $\nu=1, \dots, \kappa$, so that $\kappa \times n$ linear conditions on the coefficients $W_{i,j}$ in terms of the 
constraint functions $g_\nu$ and their derivatives can be written.
The relations (\ref{llibre2}) imply the transpositional rule (\ref{transpruleab}): indeed, by multiplying each of (\ref{llibre2}) by $\delta q_i$ and summing for $i=1, \dots, n$ one gets
\begin{equation}
	\label{llibredq}
	0=\sum\limits_{i=1}^n {\cal D}_i g_\nu\delta q_i- \sum\limits_{i,j=1}^n W_{j,i}\dfrac{\partial g_\nu}{\partial {\dot q}_j}\delta q_i = 
	\sum\limits_{i=1}^n {\cal D}_i g_\nu\delta q_i- \sum\limits_{j=1}^n \dfrac{\partial g_\nu}{\partial {\dot q}_j}\left( \delta {\dot q}_j - \dfrac{d}{dt}(\delta q_j)\right),  
	\quad \nu=1, \dots, \kappa
\end{equation}
(the last equality is due to (\ref{tra})) and we meet exactly (\ref{transpruleab}) (we deduce that assumptions $(A)$ and $(B)$ are implicitly in force). The reversed implication is not true, namely the identities (\ref{llibredq}) do not entail (\ref{llibre2}) (the vanishing of the sum does not imply the vanishing of each term).
In other words, the transpositional rule (\ref{transpruleab}) writes only $n$ conditions for the entries $W_{i,j}$, whereas (\ref{llibre2}) provides $\kappa \times n$ conditions. However, conditions (\ref{llibre2}) are not inferable from the transpositional rule, but they are additional hypotheses, whose meaning is unclear to us.

\subsubsection{Transpositional rule in the explicit form and assumption $(AB)$}

\noindent
We consider here the transpositional relation (\ref{transpruleexpl}), available when the explicit expressions (\ref{constrexpl}) of the constraints are given: the assumptions $(A)$ and $(B)$ correspond to (\ref{dcdip}) and (\ref{dvdip}) respectively and the relation (\ref{transpruleexpl}) is
\begin{equation}
\label{transpruleexplab}
\delta {\dot q}_{m+\nu}-\dfrac{d}{dt}\left( \delta q_{m+\nu}\right)
=\sum\limits_{r=1}^m \dfrac{\partial \alpha_\nu}{\partial {\dot q}_r}\left(\delta {\dot q}_r - \dfrac{d}{dt}\delta q_r\right)
-\sum\limits_{r=1}^m \left( {\cal D}_r \alpha_\nu  
-\sum\limits_{\mu=1}^\kappa \dfrac{\partial \alpha_\nu}{\partial q_{m+\mu}}
\dfrac{\partial \alpha_\mu}{\partial {\dot q}_r}\right)\delta q_r, \qquad \nu=1, \dots, \kappa
\end{equation}
with $\alpha_\nu=\alpha_\nu(q_1, \dots, q_n, {\dot q}_1, \dots, {\dot q}_m,t)$.
We examine now the latter identity assuming $(C_1)$ formulated in (\ref{c1}), that is the commutation relations (\ref{cr}) hold for the independent variables $q_r$, $r=1, \dots, m$. In this case the relation (\ref{transpruleexplab}) reduces to
\begin{equation}
	\label{treabc1}
	\delta {\dot q}_{m+\nu}-\dfrac{d}{dt}\left( \delta q_{m+\nu}\right)
	=
	-\sum\limits_{r=1}^m \left( {\cal D}_r \alpha_\nu  
	-\sum\limits_{\mu=1}^\kappa \dfrac{\partial \alpha_\nu}{\partial q_{m+\mu}}
	\dfrac{\partial \alpha_\mu}{\partial {\dot q}_r}\right)\delta q_r, \qquad \nu=1, \dots, \kappa.
\end{equation}
The combination of assumptions
$(A)$, $(B)$ and $(C_1)$ (dating back to the Hamiltonian--Suslov principle \cite{neimark} and frequently adopted, as in \cite{rumy2000}) simplifies (\ref{transprule}) to (\ref{treabc1}) (considered among others by \cite{papa}) and it provides the following
\begin{propr}
Assume that $(AB)$. If the constraints are in the explicit form (\ref{constrexpl}) and if $(C_1)$ (see (\ref{c1})
holds for the independent variables, Then $\delta {\dot q}_{m+\nu}=\dfrac{d}{dt}\left( \delta q_{m+\nu}\right)$ if and only if $\sum\limits_{i=1}^n {\cal D}_i g_\nu \delta q_i=0$.
\end{propr}

\noindent
{\bf Proof}. It suffices to recall (see the third formula in (\ref{formexpl})) that the right side of (\ref{treabc1}) is $-\sum\limits_{i=1}^n {\cal D}_i g_\nu \delta q_i$ $\quad \square$.

\begin{cor}
Assume $(A)$ and $(B)$: for a system with explicit constraints (\ref{constrexpl}) and independent variables $q_r$, $r=1,\dots, m$ verifying $(C_1)$, the commutation relations (\ref{cr}) are fulfilled also by the dependent variables $q_m+\nu$, $\nu=1, \dots, \kappa$, if and only if (\ref{derlagrsomma0}) for any $\nu=1, \dots, \kappa$. 
\end{cor}

\noindent
The commutation $(C_1)$ for independent variables does facilitate the question of linking the global commutation $(C_0)$ with the property $D\delta {\bf q}={\bf 0}$. The commutation of $d/dt$ and $\delta$ for all variables (that is $(C_0)$) is the point of view introduced by Volterra and Hamel (as we read in \cite{neimark}) and it is supported by more recent treatises, as \cite{pars}. The partial commutation $(C_1)$ only for independent variables was introduced by Suslov anf by  Levi--Civita and Amaldi. In spite of its simplifying contribution to the problem, $(C_1)$ has to be considered as a hypothesis (to our knowledge), in the absence of which the results (as the one stated in Proposition 2) are weaker.

%%%%%%%%%%%%%%%%%%%%%%%%%%%%%%%%%%%%%%%%%%%%%%%%%%%%%%%%%%%%%%%%%%%%%%%%%%%%%%%%%%%%%

\section{Some categories of nonholonomic constraints}

\noindent
We focus on analyzing special kinds of nonholonomic constraints in mechanical systems. These constraints, which limit motion in complex and non-integrable ways, will be examined across their two primary categories: those that are linear and those that are nonlinear with respect to the kinetic variables (generalized velocities). Distinguishing between these formulations is crucial for understanding the diverse challenges they pose in the dynamic description of systems.

\subsection{Linear kinematic constraints}

\noindent
We consider the class of constraints (\ref{glin}) where the function $g_\nu$ is linear with respect to the generalized velocities.
The transpositional rule (\ref{transprule}) takes the form 
\begin{equation}
	\label{transprellin}
	\begin{array}{l}
\overbrace{
\sum\limits_{i,j=1}^n 
\left( 
\dfrac{\partial a_{\nu,i}}{\partial q_j}{\dot q}_i+\dfrac{\partial b_\nu}{\partial q_j}
\right)
\delta q_j+\sum\limits_{j=1}^n a_{\nu,j}\delta {\dot q}_j}^{\delta^{(v)}g_\nu}-\dfrac{d}{dt}\overbrace{\left(\sum\limits_{i=1}^n 
a_{\nu,i} \delta q_i\right)}^{\delta^{(c)}g_\nu}
		\\
		\\
		=\sum\limits_{j=1}^n \underbrace{a_{\nu,j}}_{\frac{\partial g_\nu}{\partial {\dot q}_j}} \left(\delta {\dot q}_j - \dfrac{d}{dt}\delta q_j\right)
		-\sum\limits_{j=1}^n \underbrace{\left(
	\sum\limits_{i=1}^n  \left(
	\dfrac{\partial a_{\nu,j}}{\partial q_i}-\dfrac{\partial a_{\nu,i}}{\partial q_j} \right)
	{\dot q}_i+\dfrac{\partial a_{\nu,j}}{\partial t} - \dfrac{\partial b_\nu}{\partial q_j}
	\right)}_{D_jg_\nu}\delta q_j
	\end{array}
\end{equation}
As we already discussed in (\ref{dclin}), the ${\check{\rm C}}$etaev condition (\ref{dc}) applied to linear constraints overlaps the usual way of treating linear kinematic constraints based on the definition of virtual displacements \cite{neimark}.
Hence, linear constraints (\ref{glin}) refer to constraints verifying assumption $(A)$ and the transpositional rule (\ref{tr}) is of the kind (\ref{transprulea}) (namely the third sum on the left side of equality (\ref{transprellin}) vanishes), reducing to (\ref{dvderlagr}) if $(C_0)$ holds (the first sum on the left side of (\ref{transprellin}) vanishes).

\subsubsection{The explicit form}

\noindent
For linear constraints (\ref{glin}) the explicit relations (\ref{constrexpl}) are of the type
\begin{equation}
	\label{glinexpl}
{\dot q}_{m+\nu}=\sum\limits_{r=1}^m \xi_{\nu,r}({\bf q},t){\dot q}_r+\eta_\nu({\bf q,t}),
\qquad \nu=1, \dots, \kappa
\end{equation}
and (\ref{transpruleexpl}) takes the form
\begin{equation}
	\label{trlinexpl}
	\delta^{(v)}g_\nu - \dfrac{d}{dt}\delta^{(c)}g_\nu=
	\delta {\dot q}_{m+\nu}-\dfrac{d}{dt}\left( \delta q_{m+\nu}\right)
	-\sum\limits_{r=1}^m \xi_{\nu,r}\left(\delta {\dot q}_r - \dfrac{d}{dt}\delta q_r\right)
-\sum_{r,s=1}^m \beta_{sr}^{m+\nu}{\dot q}_s \delta q_r
-\sum\limits_{r=1}^m \gamma_r^\nu \delta q_r
\end{equation}	
where
\begin{eqnarray}
\label{beta}
\beta_{sr}^{m+\nu}&=&
\dfrac{\partial \xi_{\nu,s}}{\partial q_r}-
\dfrac{\partial \xi_{\nu,r}}{\partial q_s}
+\sum\limits_{\mu=1}^\kappa \left(
\dfrac{\partial \xi_{\nu,s}}{\partial q_{m+\mu}}\xi_{\mu,r}-
\dfrac{\partial \xi_{\nu,r}}{\partial q_{m+\mu}}\xi_{\mu,s}\right)\\
\label{gamma}
\gamma_r^\nu&=&
\dfrac{\partial \eta_\nu}{\partial q_r}-\dfrac{\partial \xi_{\nu,r}}{\partial t}
	+\sum\limits_{\mu=1}^\kappa 
	\left(
	\dfrac{\partial \eta_\nu}{\partial q_{m+\mu}}\xi_{\mu,r}-
	\dfrac{\partial \xi_{\nu,r}}{\partial q_{m+\mu}}\eta_r
	\right)
\end{eqnarray}

\noindent
The assumptions $(A)$ and $(B)$ (namely $\delta^{(c)}g_\nu=0$ and $\delta^{(v)}g_\nu=0$, respectively) are expressed in the present context (\ref{glinexpl}) by (see(\ref{dcdip}) and (\ref{dvdip}))
$$
\begin{array}{ll}	
\delta q_{m+\nu}=\sum\limits_{r=1}^m \xi_{\nu,r}\delta q_r & (A)\\
\\
\delta {\dot q}_{m+\nu}=
\sum\limits_{i=1}^n \left(
\dfrac{\partial \xi_{\nu,r}}{\partial q_i}+\dfrac{\partial \eta_\nu}{\partial q_i}\right) 
\delta q_i+
\sum\limits_{r=1}^m \dfrac{\xi_{\nu,r}}{\partial {\dot q}_r}\delta {\dot q}_r & (B)
\end{array}
$$
and we recognise in $(A)$ the standard way to express the dependent variations $q_{m+\nu}$, $\nu=1, \dots, \kappa$ in terms of the independent variations $q_r$, $r=1, \dots, m$ (see for instance \cite{neimark}).  If the commutation $(C1)$ (see (\ref{c1})) is assumed for independent variables, then the first sum on the left side of (\ref{trlinexpl}) is zero.

\noindent
In the case (with respect to (\ref{glinexpl})) 
\begin{equation}
	\label{linaut}
\xi_{\nu,j}=\xi_{\nu,r}({\bf q}), \qquad \eta_\nu=0\;\; for \;each \;\;r=1,\dots, m\;\; and \;\;\nu=1, \dots, \kappa
\end{equation}
all the coefficients $\gamma_r^\nu$ in (\ref{gamma}) vanish and the last sum in (\ref{trlinexpl}) is zero.

\begin{rem}
According to \cite{neimark}, transpositional relations used in investigations concerning the case (\ref{linaut}) are formulae (6.1), Chapter III of the same text, corresponding in our notation to
$$
\delta {\dot q}_{m+\nu}-\dfrac{d}{dt}\left( \delta q_{m+\nu}\right)
=\sum_{r,s=1}^m \beta_{sr}^{m+\nu}{\dot q}_s \delta q_r
$$
which can be applied only if one assumes (explicitly or tacitly) $(A)$, $(B)$ and $(C_1)$. 
\end{rem}

\subsubsection{Linear homogeneous constraints}
Linear autonomous constraints 
	\begin{equation}
		\label{glinaut}
		g_\nu =\sum\limits_{j=1}^n a_{\nu,j}({\bf q}){\dot q}_j=0
	\end{equation}	
which assume (in relation to (\ref{glin})) $a_{\nu,i}=a_{\nu,i}({\bf q})$, $b_\nu =0$ are largely considered and present a wide number of applications, starting from the primary examples of nonholonomic systems.
The relation (\ref{transprellin}) reduces to
\begin{equation}
	\label{transprellinaut}
	\begin{array}{l}
		\overbrace{
			\sum\limits_{i,j=1}^n 
			\dfrac{\partial a_{\nu,i}}{\partial q_j}{\dot q}_i
			\delta q_j+\sum\limits_{j=1}^n a_{\nu,j}\delta {\dot q}_j}^{\delta^{(v)}g_\nu}-\dfrac{d}{dt}\overbrace{\left(\sum\limits_{i=1}^n 
			a_{\nu,i} \delta q_i\right)}^{\delta^{(c)}g_\nu}
		\\
		\\
		=\sum\limits_{j=1}^n \underbrace{a_{\nu,j}}_{\frac{\partial g_\nu}{\partial {\dot q}_j}} \left(\delta {\dot q}_j - \dfrac{d}{dt}\delta q_j\right)
		- \sum\limits_{j=1}^n \underbrace{
			\sum\limits_{i=1}^n  \left(
			\dfrac{\partial a_{\nu,j}}{\partial q_i}-\dfrac{\partial a_{\nu,i}}{\partial q_j} \right)
			{\dot q}_i}_{D_jg_\nu}\delta q_j
	\end{array}
\end{equation}
Given that (\ref{glinaut}) is part of linear constraints, we presume that $(A)$ is satisfied, so that the total derivative $\frac{d}{dt}$ is zero in (\ref{transprellinaut}). The condition (\ref{derlagrsomma0}) is in the present case
\begin{equation}
	\label{derlagrsomma0aut}
	\sum\limits_{i,j=1}^n  \left(
	\dfrac{\partial a_{\nu,j}}{\partial q_i}-\dfrac{\partial a_{\nu,i}}{\partial q_j} \right)
	{\dot q}_i\delta q_j=0
\end{equation}
and applying (\ref{dvderlagr}) of Proposition 1 we see that if (\ref{derlagrsomma0aut}) is verified, also $(B)$ is true, in case of commutation $(C_0)$. On the other hand, if both $(A)$ and $(B)$ 
hold, then (\ref{derlagrsomma0aut}) imposes to the transpositional rule (\ref{tr}) the condition (see (\ref{abc1}))
\begin{equation}
\label{abc0linaut}
\sum\limits_{j=1}^n a_{\nu,j} \left(\delta {\dot q}_j - \dfrac{d}{dt}\delta q_j\right)
=0
\end{equation}

\noindent
We focus now on the category (\ref{glinaut}) joined with the assignment $(A)$ (consistently with the linear case), wondering whether the identity (\ref{derlagrsomma0aut}) is verified. 
We outline the problem in a vector--matrix notation by writing
\begin{equation}
\label{conj}
\left\{
\begin{array}{l}
{\bf a}_\nu \cdot {\dot {\bf q}}=0\\
\\
{\bf a}_\nu \cdot \delta {\bf q}=0
\end{array}
\right.
\qquad \Longrightarrow \qquad {\dot {\bf q}}^T\left(
\left(\dfrac{\partial {\bf a}_\nu}{\partial {\bf q}}\right)^T-
\dfrac{\partial {\bf a}_\nu}{\partial {\bf q}}\right) \delta {\bf q}\underbrace{=}_{?}0
\end{equation}
where ${\dot {\bf q}}=({\dot q}_1, \dots, {\dot q}_n)^T$, $\delta {\bf q}=(\delta q_1, \dots, \delta q_n)^T$ are (previously introduced) $n\times 1$ vectors and 
$$
{\bf a}_\nu=(a_{\nu,1}, \dots, a_{\nu,n})^T, \qquad 
\dfrac{\partial {\bf a}_\nu}{\partial {\bf q}}=
\left( 
\dfrac{\partial a_{\nu,i}}{\partial q_j}
\right)_{i,j=1, \dots, n}
$$
($\nu$ is fixed among $1,\dots, \kappa$) are respectively a $n \times 1$ vector and a $n\times n$ matrix.
The two conditions to the left of the arrow in (\ref{conj}) correspond respectively to the constraint (\ref{glinaut}) and to condition $(A)$ of (\ref{abvett}). The two vectors ${\dot {\bf q}}$ and $\delta {\bf q}$ are placed into the $(n-1)$--dimensional space $\langle {\bf a}_\nu\rangle^T$ formed by all the vectors orthogonal to ${\bf a}_\nu$ (notice that ${\bf a}_\nu=\frac{\partial g_\nu}{\partial {\dot {\bf q}}}$ cannot be zero, owing to  (\ref{gind})). The expression to the right of the arrow in (\ref{conj}) can be contextualised as the product of the two ${\Bbb R}^n$--vectors ${\dot {\bf q}}$ and $\delta {\bf q}$ in the bilinear form whose matrix is ${\Bbb A}=\left(\dfrac{\partial {\bf a}_\nu}{\partial {\bf q}}\right)^T-
\dfrac{\partial {\bf a}_\nu}{\partial {\bf q}}$. It is evident that ${\Bbb A}^T=-{\Bbb A}$, that is the matrix ${\Bbb A}$ is skew--symmetric. 
\begin{rem}
If two vectors ${\bf v}_1$ and ${\bf v}_2$ are parallel, then their product 
in a skew--symmetric bilinear form is zero. Indeed ${\bf v}_2=\lambda {\bf v}_1$ for some real value $\lambda\not =0$ (we exclude the trivial case of null vectors) and 
$$
{\bf v}_1^T {\Bbb A}{\bf v}_2={\bf v}_1^T {\Bbb A}(\lambda {\bf v}_1)=
\lambda {\bf v}_1^T {\Bbb A}{\bf v}_1=0
$$
(we recall that ${\bf v}_1^T {\Bbb A}{\bf v}_1=0$ for ${\Bbb A}$ skew--symmetric).
\end{rem}

\noindent
A special case concerns bidimensional systems, that is systems pertaining to two generalized variables $(q_1, q_2)$

\begin{prop}
For $n=2$ the conjecture (\ref{conj}) is true (i.~e.~(\ref{abc0linaut}) holds).
\end{prop}

\noindent
{\bf Proof}. It suffices to notice that the two conditions in (\ref{conj}), namely
$
\left\{
\begin{array}{l}
a_{\nu,1}{\dot q}_1+a_{\nu,2}{\dot q}_2=0\\
a_{\nu,1}\delta q_1+a_{\nu,2}\delta q_2=0
\end{array}
\right.
$
entails that the two vectors $({\dot q}_1, {\dot q}_2)^T$ and $(\delta q_1, \delta q_2)^T$ have the same directions. Geometrically, the dimension of the orthogonal space $\langle {\bf a}_\nu \rangle^T$ is one, so that all the vectors belonging to it are parallel. By virtue of the previous Remark the product to the right of (\ref{conj}), which for $n=2$ assumes the form 
$$
\begin{array}{c}
({\dot q}_1, {\dot q}_2)^T\\
\\
\end{array}
\left(
\begin{array}{cc} 
0 & \frac{\partial a_{\nu,2}}{\partial q_1}-\frac{\partial a_{\nu,1}}{\partial q_2}\\ 
\frac{\partial a_{\nu,1}}{\partial q_2}-\frac{\partial a_{\nu,2}}{\partial q_1}& 0
\end{array}\right) \left( \begin{array}{c} \delta q_1 \\ \delta q_2 \end{array} \right)
$$
is zero. $\quad \square$

\noindent
For $n>2$ the dimension of the orthogonal space $\langle {\bf a}_\nu\rangle^\perp$ is $n-1\geq 2$ and the skew--symmetric form is not necessarily null (in other words: the two vectors ${\dot {\bf q}}$ and $\delta {\bf q}$ which are both orthogonal to ${\bf a}_\nu$ are not necessarily parallel, since they belong to a vector space of dimension at least two).

\subsubsection{Exact constraints}

\noindent
By exact (or integrable) constraint we mean a constraint (\ref{vincg}) where the function $g_\nu$ satisfies (\ref{gint}) for some $f_\nu({\bf q},t)$. An exact constraint is linear and, with respect to (\ref{glin}), it is $a_{\nu,j}=\dfrac{\partial f_\nu}{\partial q_j}$, $b_\nu = \dfrac{\partial f_\nu}{\partial t}$. The relations
\begin{equation}
\label{rele}
\dfrac{\partial {\dot f}_\nu}{\partial {\dot q}_i}=\dfrac{\partial f_\nu}{\partial q_i},
\qquad \dfrac{d}{dt}\dfrac{\partial f_\nu}{\partial q_i}= \dfrac{\partial {\dot f}_\nu}{\partial q_i}
\end{equation}
which can be easily checked entail for the lagrangian derivative (\ref{derlagr})
\begin{equation}
\label{derlagre}
D_i{\dot f}_\nu=\dfrac{d}{dt}\dfrac{\partial {\dot f}_\nu}{\partial {\dot q}_i}-
\dfrac{\partial {\dot f}_\nu}{\partial q_i}=
\dfrac{d}{dt}\dfrac{\partial f_\nu}{\partial q_i}-
\dfrac{\partial {\dot f}_\nu}{\partial q_j}=
\dfrac{\partial {\dot f}_\nu}{\partial q_i}-
\dfrac{\partial {\dot f}_\nu}{\partial q_i}=0
\end{equation}
for each $i=1, \dots, n$ (we assume for $f_\nu$ continuous second partial derivatives so that the order of partial derivatives can be exchanged). Therefore, the transpositional rule (\ref{transpruleg}) is
\begin{equation}
	\label{transprulee}
\delta^{(v)} {\dot f}_\nu -\dfrac{d}{dt}\left( \delta^{(c)}{\dot f}_\nu \right) =
\sum\limits_{j=1}^n 
\dfrac{\partial f_\nu}{\partial q_j}\left(\delta {\dot q}_j - \dfrac{d}{dt}\delta q_j\right)
\end{equation}
We already remarked (equation (\ref{dce})) that the $\check{\rm C}$etaev condition for $g_\nu$ coincides with the standard definition of virtual displacements for the corresponding holonomic constraint $f_\nu({\bf q}, t)=0$. Hence it is expected to have as certain  $\delta^{(c)}({\dot f}_\nu)=0$ (assumption $(A)$), otherwise we would contradict a consolidated point in the theory of holonomic systems.
Thus, exact constraints satisfy the transpositional rule in the form (\ref{transprulead0}) and if the commutation relation $(C_0)$ holds then also $\delta^{(v)}{\dot f}_\nu=0$ (assumption $(B)$).
The validity of the commutation relations (\ref{cr}) is usually 
accepted for holonomic systems, so that in the case of exact constraints each of the three elements in (\ref{transprulee}) (that is the two terms with $\delta^(v)$, $\delta^{(c)}$ and the summation) are zero.

\begin{rem}
The necessity of assuming the commutation relations (\ref{cr}) for exact constraints can be also argumented as follows. If we define consistently with (\ref{dv}) 
\begin{equation}
\label{dvol}
\delta^{(v)}f_\nu ({\bf q,t})=\sum\limits_{j=1}^n \dfrac{\partial f_\nu}{\partial q_j} \delta q_j
\end{equation}
(we incidentally notice that $\delta^{(c)}f_\nu=0$, on the contrary, does not make sense), 
one has, by virtue of (\ref{rele}),
	$$
	\dfrac{d}{dt} (\delta^{(v)} f_\nu) = \sum\limits_{j=1}^n\left(
	\dfrac{d}{dt}\left(\dfrac{\partial f_\nu}{\partial q_j}\right)\delta q_j
	+\dfrac{\partial f_\nu}{\partial q_j}\dfrac{d}{dt}(\delta q_j)\right)=
	 \sum\limits_{j=1}^n
	 \left(
	\dfrac{\partial {\dot f}_\nu}{\partial q_j}\delta q_j
	+\dfrac{\partial {\dot f}_\nu}{\partial {\dot q}_j}\dfrac{d}{dt}(\delta q_j)\right)
$$
while
$$
\delta^{(v)} {\dot f}_\nu = \sum\limits_{j=1}^n
\left(
\dfrac{\partial {\dot f}_\nu}{\partial q_j}\delta q_j
+\dfrac{\partial {\dot f}_\nu}{\partial {\dot q}_j}(\delta {\dot q}_j)\right)
$$
so that $\dfrac{d}{dt} (\delta^{(v)} f_\nu) =\delta^{(v)} {\dot f}_\nu$ only if (\ref{cr}) are true (in other words, the commutation relations (\ref{cr}) imply the commutation between $d/dt$ and $\delta^{(v)}$). 
Now, assume for the system the constraint $f_\nu ({\bf q},t)=0$: ${\dot f}_\nu ({\bf q}, {\dot {\bf q}},t)=0$ is a constraint, too, but the congruence between $\delta^{(v)} f_\nu=0$ and  $\delta^{(v)} {\dot f}_\nu=0$ is guaranteed only if the commutation relations are satisfied.
\end{rem}

\noindent
We summarize the argument by means of the following 
\begin{propr}
If the commutation relations (\ref{cr}) are satisfied, then the operations $\dfrac{d}{dt}$ and $\delta^{(c)}$ commute for any function $f_\nu({\bf q},t)$ (see (\ref{dvol})). Conversely, 
the simultaneous validity of $\delta^{(v)} f_\nu=0$ and  $\delta^{(v)} {\dot f}_\nu=0$ 
(demanded by the same requirement of the two constraints $f_\nu=0$, ${\dot f}_\nu=0$) implies the fulfillment of the commutation relations (\ref{cr}).
\end{propr}

\subsubsection{Integrable constraints via an integrating factor}

\noindent
Let us examine an intermediate class between exact constaints and linear constraints, discussed in \cite{flan} and revealing interesting contacts with the transpositional rule (\ref{tr}).
We say that the constraint $g_\nu$ is integrable via an integrating factor $\phi_\nu({\bf q}, t)$ if the function $g_\nu \phi_\nu$ is exact, namely
\begin{equation}
\label{gif}
\phi_\nu ({\bf q},t)g_\nu({\bf q}, {\dot {\bf q}},t)=\dfrac{d}{dt}f_\nu ({\bf q},t).
\end{equation}
As an example, consider for $n=2$ the kinematic constraint $g_1(q_1, q_2, {\dot q}_1, {\dot q}_2)=q_2{\dot q}_1-q_1{\dot q}_2=0$: the function $g_1$ is not exact but 
$\dfrac{1}{q_2^2}({\dot q}_1 q_2 -{\dot q}_2 q_1)=\dfrac{d}{dt}(q_1/q_2)$, so that 
(\ref{gif}) is satisfied with the integrating factor $\phi_1=1/q_2^2$ and $f_1=q_1/q_2$.

\noindent
The function $g_\nu$ appearing in (\ref{gif}) is necessarily linear in the generalized velocities ${\dot {\bf q}}$ and with respect to (\ref{glin}) it must be
\begin{equation}
	\label{relif}
	\phi_\nu a_{\nu,j}=\dfrac{\partial f_\nu}{\partial q_j}, \qquad \phi_\nu b_\nu=\dfrac{\partial f_\nu}{\partial t}
\end{equation}
Conversely, the following closure conditions are necessary for (\ref{gif}) to be true:
\begin{equation}
	\label{chiusaif}
	\left\{
	\begin{array}{ll}
		\phi_\nu \left( 
		\dfrac{\partial a_{\nu,i}}{\partial q_j}-\dfrac{\partial a_{\nu,j}}{\partial q_i}
		\right)
		=a_{\nu, j} \dfrac{\partial \phi_\nu}{\partial q_i}-
		a_{\nu, i} \dfrac{\partial \phi_\nu}{\partial q_j}&
		\\
		&i,j=1, \dots, n
		\\ 
		\phi_\nu \left( \dfrac{\partial b_\nu}{\partial q_j}-\dfrac{\partial a_{\nu,j}}{\partial t}\right)
		=
		a_{\nu, j} \dfrac{\partial \phi_\nu}{\partial t}-
		b_\nu \dfrac{\partial \phi_\nu}{\partial q_j}&
	\end{array}
	\right.
\end{equation}
and they are sufficient in any simply connected open subset (by the Poincar\'e Lemma). Obviously (\ref{gif})
extends (\ref{gint}), which corresponds to $\phi_\nu=1$: in this case 
the right-hand sides of equalities (\ref{chiusaif}) vanish and they replicate the 
standard condition of closure for the differential form $\sum\limits_{j=1}^n a_{\nu,j} dq_j+b_\nu dt$.

\begin{rem}
It should appear appropriate to consider only $\phi_\nu\not =0$, so that the condition $\phi_\nu g_\nu=0$ is equivalent to the given constraint $g_\nu=0$. Actually, if the function $\phi_\nu$ is not identically zero, the points where $\phi_\nu=0$ may add constraint configurations  extraneous to $g_\nu=0$.
Nevertheless, we remark that when $\phi_\nu=0$ for some value ${\bf q}$ and $t$, by
(\ref{chiusaif}) we deduce $a_{\nu, j} \frac{\partial \phi_\nu}{\partial q_i}=
a_{\nu, i} \frac{\partial \phi_\nu}{\partial q_j}$, 
$a_{\nu, j} \frac{\partial \phi_\nu}{\partial t}=b_\nu \frac{\partial \phi_\nu}{\partial q_j}$ and this implies that the coefficients of the two constraints $g_\nu=\sum\limits_{j=1}^n 
a_{\nu,j}{\dot q}_j+b_\nu=0$ and ${\dot \phi}_\nu=
\frac{\partial \phi_\nu}{\partial q_j}{\dot q}_j+\frac{\partial \phi_{\nu}}{\partial t}=0$
are proportional, so that they are fulfilled by the same $n$--uples ${\dot {\bf q}}$; in this way $\phi_\nu=0$ does not add further constraint configurations, that is the restrictions $\phi_\nu g_\nu =0$ and $g_\nu=0$ are equivalent. 
\end{rem}

\begin{propr}
	Assume $g_\nu$ of the type (\ref{gif}) for some integrating factor $\phi_\nu$. Then, the Lagrangian derivative of $g_\nu$ satisfies the identity
	\begin{equation}
		\label{derlagrif}
		\phi_\nu {\cal D}_j g_\nu= g_\nu \dfrac{\partial \phi_\nu}{\partial q_j}-{\dot \phi}_\nu
		\dfrac{\partial g_\nu}{\partial {\dot q}_j} .
	\end{equation}
\end{propr}

\noindent
{\bf Proof}: it suffices to recall (\ref{derlagre}) in order to write
$$
0= {\cal D}_j{\dot f}_\nu= {\cal D}_j(\phi_\nu g_\nu)=
\dfrac{d}{dt}\left(\phi_\nu \dfrac{\partial g_\nu}{\partial {\dot q}_j}\right)
-\phi_\nu \dfrac{\partial g_\nu}{\partial q_j}
-g_\nu \dfrac{\partial \Phi_\nu}{\partial q_j}=
\overbrace{\phi_\nu \dfrac{d}{dt}\left( \dfrac{\partial g_\nu}{\partial {\dot q}_j}\right)
-\phi_\nu \dfrac{\partial g_\nu}{\partial q_j}}^{=\phi_\nu {\cal D}_j g_\nu}
-{\dot \phi}_\nu \dfrac{\partial g_\nu}{\partial {\dot q}_j} 
-g_\nu \dfrac{\partial \Phi_\nu}{\partial q_j}.
\quad \square
$$
\begin{cor}
	For the constraint $g_\nu=0$, where $g_\nu$ is of the type (\ref{gif}), the lagrangian derivative verifies
	\begin{equation}
		\label{derlagrgif}
		\phi_\nu {\cal D}_j g_\nu=-{\dot \phi}_\nu
		\dfrac{\partial g_\nu}{\partial {\dot q}_j}.
	\end{equation}
\end{cor}

\begin{rem}
	When $\phi_\nu$ is a non--zero constant then $g_\nu$ is exact and (\ref{derlagrgif}) reproduces (\ref{derlagre}).
\end{rem}

\noindent
Multiplying (\ref{derlagrgif}) by $\delta q_j$ and summing with respect to $j$, we get
\begin{equation}
\label{derlagrsif}
\phi_\nu \sum\limits_{j=1}^n {\cal D}_j g_\nu \delta q_j = 
-{\dot \phi}_\nu \sum\limits_{j=1}^n \dfrac{\partial g_\nu}{\partial {\dot q}_j} \delta q_j
\end{equation}
whence the following 

\begin{prop}
Let $g_\nu=0$ be a constraint of the type (\ref{gif}) for some integrating factor $\phi_\nu$ and some function $f_\nu$. If the displacements $\delta q_j$ satisfy assumption $(A)$ (see (\ref{abc0})), then condition (\ref{derlagrsomma0}) is satisfied, wherever the integrating factor $\phi_\nu$ is not zero.  
\end{prop}

\begin{rem}
	The ${\check {\rm C}}$etaev condition $\delta^{(c)}g_\nu=0$ (condition $(A)$) for a constraint of the type (\ref{gif}) verifies $\phi_\nu \delta^{(c)}g_\nu = \sum\limits_{i=1}^n \frac{\delta f_\nu}{\delta q_i}\delta q_i$, which extends (\ref{dce}.)
\end{rem}

\noindent
Coming back to the exmple at the beginning of the Paragraph, we see that the two lagrangian derivatives ${\cal D}_1 g_1=2{\dot q}_2$, ${\cal D}_2 g_1=-2{\dot q}_1$ do not vanish separately (actually the constraint is not exact), but the sum (\ref{derlagrsomma0}) which corresponds to
$2{\dot q}_2 \delta q_1-2{\dot q}_1 \delta q_2$ vanishes along the constraint $g_1=0$, provided that the 
$\check{\rm C}$etaev condition $\frac{\partial g_1}{\partial {\dot q}_1}\delta q_1 +
\frac{\partial g_1}{\partial {\dot q}_2}\delta q_2=q_2 \delta q_1-q_1 \delta q_2 =0$ holds. Indeed:
$$
2{\dot q}_2 \delta q_1-2{\dot q}_1 \delta q_2\overbrace{=}^{\delta q_2=(q_2/q_1)\delta q_1}
\dfrac{2}{q_1}(\overbrace{{\dot q}_2q_1-{\dot q}_1q_2}^{=g_1})\delta q_1=0.
$$

\noindent
The effect of the transpositional rule (\ref{transprule}) on integrable constraints via an integrating factor is the same as for exact contraints:  even though the lagrangian derivatives are not null separately, their combination (\ref{derlagrsomma0}) vanishes if the constraint fulfills (\ref{dc}). 
Assuming $(A)$, the rule (\ref{transprule}) to be applied for constraints (\ref{gif}) is (\ref{transprulead0}).

\noindent
It is known that the lagrangian derivative of the function $g_\nu ({\bf q}, \dot {\bf q},t)$ is zero if and only if (at least locally) the function $g_\nu$ is the total derivative of a function $f ({\bf q},t)$: $D_j g_\nu =0$ $\leftrightarrow$ $g_\nu ={\dot f}$.
It follows that an integrable constraint (\ref{gint}) verifies condition (\ref{derlagrsomma0}). However, the integrability of $g_\nu$ is not a necessary condition for (\ref{derlagrsomma0}),since the existence of an integrating factor guarantees such a property, as the latter Proposition shows.

\noindent
An interesting (and apparently not trivial) problem really concerns the ``inverse'' statement: we may wonder whether the fulfillment of (\ref{derlagrsomma0}) in the class of displacements verifying condition $(A)$ (see (\ref{abc0})) entails that the constraint is necessarily of the type (\ref{gif}). If true, the category (\ref{gif}) exhausts the constraints for which the transpositional rule in the reduced form (\ref{transprulead0}) can be applied.

\subsection{Nonlinear kinematic constraints}

\noindent
Within the category of nonholonomic constraints, a critical distinction lies between constraints that are linear and nonlinear in velocities. Most literature and classic examples of nonholonomic systems (like the rolling disk or rolling sphere) focus on constraints that are linear or affine in generalized velocities). In the nonlinear case, compatibility of virtual variations with nonlinear constraints, ideality of the constraints, conditions like ${\check {\rm C}}$etaev's must be applied with care and are less straightforward to interpret. Furthermore, variational principles or their extensions might not be directly applicable or may require substantial modifications in the presence of nonlinear constraints. 

\noindent
The realization of a mechanical nonholonomic model with nonlinear constraints dates back to \cite{hamel} and  the system considered (Appell--Hamel machine) has been largely considered in literature (see \cite{liromp} and the quoted literature).

\noindent
In spite of the problem of a physical implementation of a nonlinear nonholonomic model, from a theoretical point of view we can imagine simple and natural conditions corresponding to nonlinear restrictions, as for instance parallelism or orthogonality of the velocities of more points, magnitude of the velocity assigned, or the nonholonomic pendulum presented in \cite{benentipendulum}.

\noindent
We select two categories of nonlinear constraints, which contain many significant models and examples: homogeneous constraints and constraints independent of spatial coordinates.

\subsubsection{Homogeneous constraints}

\noindent
The function $g_\nu({\bf q}, {\dot {\bf q}},t)$ is positive homogeneous of degree $p\in {\Bbb R}$ with respect to the variables ${\dot {\bf q}}$ if 
\begin{equation}
\label{hom}
g_\nu({\bf q}, \lambda {\dot {\bf q}},t)=\lambda^p g_\nu({\bf q}, {\dot {\bf q}},t)\qquad \forall \lambda >0.
\end{equation}

\noindent
Several nonlinear (with respect to ${\dot {\bf q}}$) restrictions comply with homogeneous conditions as for instance parallel or orthogonal velocities of two or more points. The model presented in \cite{benentipendulum}, equipped by the description of a physical realization, shows homogeneous constraints of degree two (nonholonomic pendulum).

\noindent
We recall the following 
\begin{teo}
(Euler's Theorem on homogeneous functions). The function $g_\nu$ is positive homogeneous w.~r.~t.~${\dot {\bf q}}$ of degree $p$ if and only if 
\begin{equation}
	\label{eulero}
	\sum\limits_{i=1}^n {\dot q}_i \dfrac{\partial g_\nu}{\partial {\dot q}_i}=p g_\nu.
\end{equation}
\end{teo}

\noindent
From (\ref{eulero}) we deduce that the constraint condition $g_\nu=0$ can be expressed as
\begin{equation}
\label{eulero0}
\sum\limits_{i=1}^n {\dot q}_i \dfrac{\partial g_\nu}{\partial {\dot q}_i}=0.
\end{equation}
The formal affinity of the previous relation with condition $(A)$ (see relation (\ref{abc0}))
leads many authors to approve this condition for the case of homogeneous constraints (see for instance \cite{flan}).
\begin{rem}
In terms of explicit functions (\ref{constrexpl}), in case of homogeneous constraints (\ref{hom}) the functions ${\dot q}_{m+\nu}=\alpha_\nu$ are homogeneous with respect to ${\dot q}_m$, $m=1, \dots, r$, of degree $1$, hence they verify (see (\ref{eulero}))
$$
{\dot q}_{m+\nu}=\alpha_\nu =\sum\limits_{r=1}^n \dfrac{\partial \alpha_\nu}{\partial {\dot q}_r}{\dot q}_r
$$
and the affinity of such a condition with (\ref{dcdip}) supports the adoption of the ${\check{\rm C}}$etaev condition $(A)$ for homogeneous constraints.
\end{rem}

\noindent
The transpositional rule pertaining to condition $(A)$ is (\ref{transprulea}) and the coupling of $(A)$ and (\ref{eulero}) gives
\begin{equation}
	\label{condhom}
	\left\{
	\begin{array}{l}
	\dfrac{\partial g_\nu}{\partial {\dot {\bf q}}} \cdot {\dot {\bf q}}=0\\
		\\
	\dfrac{\partial g_\nu}{\partial {\dot {\bf q}}} \cdot \delta {\bf q}=0
\end{array}
\right.
\end{equation} 
which places the generalized velocities ${\dot {\bf q}}$ and the displacements $\delta {\bf q}$ into the vector space $(\ref{w})$.

\noindent
Let us examine in particular polinomial homogeneous functions of degree $2$
$$
g_\nu=\sum\limits_{i,k=1}^n \gamma_{i,k}^{(\nu)}({\bf q}){\dot q}_i {\dot q}_k
$$
which encompass several examples of nonholonomic restrictions.

\noindent
Since 
$\dfrac{\partial g_\nu}{\partial {\dot q}_j}=
\sum\limits_{i=1}^n \left(\gamma_{j,i}^{(\nu)}+\gamma_{i,j}^{(\nu)}\right)
{\dot q}_i$, condition $(A)$ (corresponding to the second line in (\ref{condhom})) is
$$
\delta^{(c)}g_\nu = \sum\limits_{i=1}^n \left(\gamma_{j,i}^{(\nu)}+\gamma_{i,j}^{(\nu)}\right)
{\dot q}_i\delta q_j =0.
$$
and (\ref{transprulea}) is
\begin{equation}
	\label{transphom}
\begin{array}{l}
\overbrace{\sum\limits_{i,j,k=1}^n \frac{\partial \gamma_{i,k}^{(\nu)}}{\partial q_j}{\dot q}_i {\dot q}_k \delta q_j+\sum\limits_{i,j=1}^n \left(\gamma_{i,j}^{(\nu)}+\gamma_{j,i}^{(\nu)}\right){\dot q}_i\delta {\dot q}_j}^{\delta^{(v)}g_\nu}=
\sum\limits_{i,j=1}^n \left(\gamma_{i,j}^{(\nu)}+\gamma_{j,i}^{(\nu)}\right){\dot q}_i
\left(\delta {\dot q}_j -\frac{d}{dt}(\delta q_j)\right)\\
\\
-\sum\limits_{j=1}^n \underbrace{\left( \sum\limits_{i=1}^n \left(\gamma_{i,j}^{(\nu)}+\gamma_{j,i}^{(\nu)}\right){\q2dot\limits^{..}}_i +
\sum\limits_{i,k=1}^n \left( 
\frac{\partial \gamma_{i,j}^{(\nu)}}{\partial q_k}
+\frac{\partial \gamma_{j,i}^{(\nu)}}{\partial q_k}
-\frac{\partial \gamma_{i,k}^{(\nu)}}{\partial q_j}
\right) {\dot q}_i {\dot q}_k\right)}_{D_j g_\nu} \delta q_j
\end{array}
\end{equation}
which plays different roles according to the assumed hypotheses, which can be 
\begin{itemize}
	\item{} if the commutation $(C_0)$ is supposed, then (\ref{transphom}) corresponds to (\ref{dvderlagr}) and we expect $\delta^{(v)}g_\nu \not =0$,
\item{} if both $(A)$ and $(B)$ are assumed, then (\ref{transphom}) corresponds to (\ref{transpruleab}) and the commutation $(C_0)$ cannot be satisifed.  
\end{itemize}
The previous statements are motivated by the invalidity of condition (\ref{derlagrsomma0}), except for particular cases: actually, based on (\ref{zero2}) of Proposition 3 it should be
$$
\sum\limits_{i=1}^n \left(\gamma_{i,j}^{(\nu)}+\gamma_{j,i}^{(\nu)}\right){\q2dot\limits^{..}}_i +
\sum\limits_{i,k=1}^n \left( 
\frac{\partial \gamma_{i,j}^{(\nu)}}{\partial q_k}
+\frac{\partial \gamma_{j,i}^{(\nu)}}{\partial q_k}
-\frac{\partial \gamma_{i,k}^{(\nu)}}{\partial q_j}
\right) {\dot q}_i {\dot q}_k = 
\sum\limits_{\mu=1}^\nu \sum\limits_{i=1}^n \varrho^{(\nu)}_\mu \left(\gamma_{j,i}^{(\mu)}+\gamma_{i,j}^{(\mu)}\right)
{\dot q}_i\qquad j=1, \dots, n
$$
for some functions $\varrho^{(\nu)}_\mu({\bf q}, {\dot {\bf q}})$, but this relation cannot be an identity.

\noindent
The coupling of assumptions $(A)$ and $(B)$ joined with the statement $(C)$ in (\ref{abc0})
(proposed in order to conciliate the vakonomic variational method with the D'Alembert--Lagrange nonholonomic method based ob $(A)$, see \cite{pastore}) makes (\ref{transphom}) of the form
(\ref{llibredq}), namely
$$
\sum\limits_{i,j,k=1}^n
\left( 
\left(\gamma_{i,j}^{(\nu)}+\gamma_{j,i}^{(\nu)}\right){\q2dot\limits^{..}}_i +
\left( 
\frac{\partial \gamma_{i,j}^{(\nu)}}{\partial q_k}
+\frac{\partial \gamma_{j,i}^{(\nu)}}{\partial q_k}
-\frac{\partial \gamma_{i,k}^{(\nu)}}{\partial q_j}
\right) {\dot q}_i {\dot q}_k-
W_{i,j} 
\left(\gamma_{i,k}^{(\nu)}+\gamma_{k,i}^{(\nu)}\right)
{\dot q}_k
\right)\delta q_j= 0
$$
where the displacements $\delta q_j$ must verify $(A)$.
Concerning the question of identifying the $n\times n$ coefficients $W_{i,j}$, the previous relation provides only one condition for each constraint $g_\nu$, $\nu=1, \dots, \kappa$.
In \cite{llibre} it is supposed that for each $j=1, \dots, n$ the expression enclosed by the external round brackets is zero (the motivation is in our opinion elusive). In that case, $\kappa \times n$ conditions for the coefficients $W_{i,j}$, $i,j=1, \dots, n$, can be written in terms of 
the functions $\gamma_{i,j}^{(\nu)}$, their derivatives w.~r.~t.~${\bf q}$ and in terms of the variables ${\dot {\bf q}}$ and $\bfq2dot^{..}$.

\subsubsection{Constraints depending only on velocities}

\noindent
We finally consider the special case 
\begin{equation}
	\label{noq}
	g_\nu=g_\nu ({\dot {\bf q}},t)
\end{equation}
that is the constraint function does not depend explicitly on the lagrangian coordinates ${\bf q}$.
For instance, the imposed magnitude of the velocity of a point $\sqrt{{\dot q}_1^2+{\dot q}_2^2+{\dot q}_3^2}=C(t)$, with $C(t)$ given nonnegative function, is of type (\ref{noq}).

\noindent
The variations (\ref{dcdef}), (\ref{dvdef}) and the lagrangian derivative (\ref{derlagr}) are
$$
\delta^{(c)}g_\nu = \dfrac{\partial g_\nu}{\partial {\dot {\bf q}}}\cdot \delta {\bf q}, \qquad
\delta^{(v)}g_\nu = \cancel{\dfrac{\partial g_\nu}{\partial {\bf q}}\cdot \delta {\bf q}}
+\dfrac{\partial g_\nu}{\partial {\dot {\bf q}}}\cdot \delta {\dot {\bf q}}, \qquad
D_j g_\nu=\dfrac{d}{dt} 
\left(\dfrac{\partial g_\nu}{\partial {\dot q}_j}\right) -
\cancel{\dfrac{\partial g_\nu}{\partial q_j}}
$$
and the relation (\ref{transprule}) is
\begin{equation}
\label{transprulenoq}
\dfrac{\partial g_\nu}{\partial {\dot {\bf q}}}\cdot \delta {\dot {\bf q}}-\dfrac{d}{dt}
\left(\dfrac{\partial g_\nu}{\partial {\dot {\bf q}}}\cdot \delta {\bf q}\right)=
\dfrac{\partial g_\nu}{\partial {\dot {\bf q}}}\cdot \left(\delta {\dot {\bf q}}-\dfrac{d}{dt}\delta {\bf q}\right)-\dfrac{d}{dt}
\left( 
\dfrac{\partial g_\nu}{\partial {\dot {\bf q}}}\right)\cdot \delta {\bf q}
\end{equation}
Assume $(A)$: the condition (\ref{zero2}) equivalent to (\ref{derlagrsomma0}) is 
\begin{equation}
\label{zero2noq}
\dfrac{d}{dt}\left(\dfrac{\partial g_\nu}{\partial {\dot {\bf q}}}\right)=\sum\limits_{\mu=1}^\kappa \varrho_\mu \dfrac{\partial g_\nu}{\partial {\dot {\bf q}}}
\end{equation}
where we expect $\varrho_\mu=\varrho_\mu ({\dot {\bf q}},t)$. If also $(B)$ holds, then (\ref{transprulenoq}) is 
$\frac{\partial g_\nu}{\partial {\dot {\bf q}}}\cdot \left(\delta {\dot {\bf q}}
-\frac{d}{dt}\delta {\bf q}\right)
=\frac{d}{dt} \left( \frac{\partial g_\nu}{\partial {\dot {\bf q}}}\right)\cdot \delta {\bf q}$
and commutation $(C_0)$ can hold only for constraints verifying (\ref{zero2noq}).

\begin{rem}
The condition $\dfrac{\partial g_\nu}{\partial {\dot {\bf q}}}\cdot \delta {\dot {\bf q}}=0$ is applied for general constraints (\ref{vincg}) (that is including the variables ${\bf q}$) in the so called Jourdain principle \cite{jourdain}. The question about the validity of the commutation relation for such a principle is discussed in \cite{papajourdain}.  
\end{rem}

\section{Conclusion}

\noindent
In this paper we have thoroughly examined the intricate interplay between various variational assumptions and the transpositional rule in the context of nonholonomic mechanical systems. Our analysis underscored how the ${\check {\rm C}}$etaev condition $\delta^{{(c)}}g_\nu=0$ and the first variation of constraints $\delta^{(v)}g_\nu=0$ each impose distinct, yet interconnected, restrictions on virtual displacements. We specifically characterized the necessary conditions for the commutation relations $\delta {\dot q}_j=\frac{d}{dt} \delta q_j$ to hold, clarifying their implications for the consistency of nonholonomic dynamics. The established relationship between the Lagrangian derivatives of the constraint functions and the commutativity assumptions provides a deeper insight into the underlying mathematical structure. Ultimately, this work highlights the critical role of these fundamental variational identities in unifying disparate approaches to nonholonomic mechanics, paving the way for more robust derivations of equations of motion. Future work will explore the practical application of these refined theoretical frameworks to complex mechanical systems and investigate the analytical properties of the $\varrho_\mu^{(\nu)}$ coefficients in nonlinear constraint scenarios.

\noindent
This study has provided a comprehensive investigation into the foundational variational principles governing nonholonomic mechanical systems, focusing on the transpositional rule as a bridge between different formalisms. We meticulously analyzed the roles of the ${\check {\rm C}}$etaev condition and the first variation of constraints, demonstrating their combined effect on the definition of compatible virtual displacements. A key finding was the identification of necessary conditions for the commutation of variational and time differentiation operators, which helps reconcile various approaches to nonholonomic dynamics. By linking these assumptions to the Lagrangian derivatives of the constraint functions, we've contributed to a clearer understanding of when and how different formulations of equations of motion (such as those from d'Alembert-Lagrange vs. extended Hamilton principles) are consistent. This deeper insight into the mathematical underpinnings of nonholonomic constraints opens avenues for more precise modeling and simulation. Future research could extend this analysis to specific classes of highly nonlinear nonholonomic systems, further exploring the implications of the derived necessary conditions for practical applications.


\begin{thebibliography}{10}
	
	\bibitem{arnoldkozlov} Arnol'd, V.~I.~, Kozlov, V.~V.~, Neishtadt, A.~I.(2006) \emph{Mathematical Aspects of Classical and Celestial Mechanics}, Third Editiom, E.~M.~S.~,Dynamical Systems III, Springer
	
	\bibitem{benentipendulum} Benenti, S.~(2011) The non--holonomic double pendulum, an example of non-linear non-holonomic system, {\emph Regular and Chaotic Dynamics}, {\bf 1} n.~5, 417--442.
	
	\bibitem{cronstrom} Cronstr{\" o}m, C.~(2010) On the Compatibility of Nonholonomic Systems and Related Variational Systems, {\emph Acta Physica Universitatis comenianae} Vol.~L--LI, 1 and 2, 25--36.
	
	\bibitem{cetaev} ${\check {\rm C}}$etaev, N.~G.~(1962) On the Gauss Principles, {\emph Papers on Analytical Mechanics} {\bf 323}, Science Academy.

	\bibitem{flan} Flannery M.~R.~(2011) d'Alembert--Lagrange analytical dynamics for nonholonomic systems, \emph{Journal of Mathematical Physics} {\bf 52}, 032705 1--29.
	
	\bibitem{flanneryenigma} Flannery M.~R.~(2005) The enigma of nonholonomic constraints, {\emph American Journal of Physics} {\bf 73}, 265--272.
	
	\bibitem{golub} Golub, G.~H.~, Van Loan, C.~F.~(2013)
	{\emph Matrix Computations} 4th ed.~Johns Hopkins University Press.
	
	\bibitem{hamel} Hamel, G.~(1909) {\emph Theoretische Mechanik}, Springer--Verlag, Berlin. 
	
		\bibitem{jourdain} Jourdain, P.~E.~B.~(1909) Note on an analogue of Gauss’ principle of least constraint {\emph Q.~J.~Pure Appl.~Math.~}{\bf 8L} 153-–157.
	
		\bibitem{krup4} Krupkov\'a, O.~(2009) The nonholonomic variational principle, {\emph J.~Phys.~A: Math.~Theor.} {\bf 42}, 185201. 
		
			\bibitem{lemos} Lemos, N.~A.~(2022) Complete inequivalence of nonholonomic and vakonomic mechanics, {\emph Acta Mech.~} {\bf 233}, 47--56.
	
		\bibitem{liromp} Li, S.~M.~, Berakadar, J.~(2007) On the validity of the vakonomic model and the Chetaev model for constraint dynamical systems, {\emph Reports on Mathematical Physics} {\bf 60} Issue 1, 107--115.
	
	\bibitem{llibre}  Llibre, J.~, Ram\'irez, R.~, Sadovaskaia, N.~(2014) A new approach to vakonomic mechanics, \emph{Nonlinear Dynamics} {\bf 78}, 2219--2247.
	
	\bibitem{lurie} Lurie, A.\,I.~(2002) \textit{Analytical Mechanics}, Springer-Verlag, Berlin Heidelberg.
	
		\bibitem{mei} Mei, F.~(2000), Nonholonomic mechanics, \emph{Appl.~Mech.~Rev.~} {\bf 11}, 283--305.
	
	
	\bibitem{neimark} Ne${\check {\rm i}}$mark Ju.~I.~, Fufaev N.~A.~(1972) Dynamics of Nonholonomic Systems, Providence: American Mathematical Society, Translations of Mathematical Monographs {\bf 33}.
	
	\bibitem{papa} J. G. Papastravidis, J.~G.~(1992) Time--integral Variational Principles for Nonlinear Nonholonomic Systems, J.~Appl.~Math.~{\bf 64}, 985--991.
	
	\bibitem{papajourdain} Papastavridis, J.~G.~(1992) On Jourdain's principle, \emph{International Journal of Engineering Science} {\bf 30} Issue 2, 135--140
	
		
	\bibitem{pars} Pars, L.~A.~(1965) \emph{A treatise on analytical dynamics}, \emph{London: Heinemann Educational Books Ltd.}
	
		
	\bibitem{pastore} Pastore, A.~, Giammarini, A.~, Grillo, A.~(2024) Reconciling Kozlov's vakonomic method with the traditional non-holonomic method: solution of two benchmark problems, {\emph Acta Mechanica} {\bf 235}, 2341--2379.
	
	
	
	\bibitem{ray} Ray, J.~R.~(1966) Nonholonomic constraints, \emph{Am.~J.~Phys.~} {\bf 34}, 406–-408. 

	\bibitem{rumy2000} V. V. Rumyantsev, V.~V.~, Forms of Hamilton's principle for nonholonomic systems, Mechanics, Automatic and Robotics {\bf 2}, n.~10, 1035--1048, 2000.
	
	
	
	\bibitem{rumycet} Rumyantsev, V.~V.~(1973) On the Chetaev principle (In Russian), \emph{Dokl.~Akad.~Nauk SSSR} {\bf 210}  n.~4, 787--790.
	

	
\end{thebibliography}
\end{document}